\documentclass[usenatbib]{mnras}
\usepackage[utf8]{inputenc}
\usepackage{graphicx}
\usepackage{multirow}
\usepackage{threeparttable}
\usepackage{amsmath}

\newcommand{\bb}[1]{#1}
\newcommand{\lise}{^7\textrm{Li}}
\newcommand{\lisi}{^6\textrm{Li}}
\newcommand{\ratio}{\lisi/\lise}
\newcommand{\alise}{\textrm{A}(^7\textrm{Li})}
\newcommand{\alisi}{\textrm{A}(^6\textrm{Li})}
\newcommand{\ali}{\textrm{A}(\textrm{Li})}
\newcommand{\ak}{\textrm{A}(\textrm{K})}

\newcommand{\teff}{T_{\rm{eff}}} 
\newcommand{\logg}{\log g}
\newcommand{\kms}{$\,{\rm km\,s^{-1}}$}
\newcommand{\ms}{$\,{\rm m\,s^{-1}}$} 
\newcommand{\met}{[\rm{Fe}/\rm{H}]}

\newcommand{\vr}{\delta v_{\rm{rad}}}
\newcommand{\tvr}{v_{\rm{rad}}}
\newcommand{\vsini}{v\sin(i)}
\newcommand{\svsini}{v\sin(i)}
\newcommand{\feI}{\ion{Fe}{i}}
\newcommand{\feII}{\ion{Fe}{ii}}
\newcommand{\kI}{\ion{K}{i}}
\newcommand{\liI}{\ion{Li}{i}}
\newcommand{\kline}{\kI\ 769.9\,nm}
\newcommand{\liline}{\liI\ 670.8\,nm}

\newcommand{\balder}{\textsc{balder}}
\newcommand{\multitd}{\textsc{multi3d}}
\newcommand{\sklearn}{\textit{scikit-learn}}

% ...
\newcommand{\bba}[1]{#1}

\begin{document}

\title[Non-detection of $\lisi$ in Spite plateau stars]{Non-detection of $\lisi$ in Spite plateau stars with ESPRESSO}
\author[E.~Wang, T.~Nordlander, M.~Asplund, K.~Lind, Y.~Zhou, H.~Reggiani]{Ella~Xi~Wang$^{1, 2}$\thanks{Email: ellawang@mso.anu.edu.au}, 
Thomas~Nordlander$^{1, 2}$, 
Martin~Asplund,  
Karin~Lind$^{3}$, \newauthor
Yixiao~Zhou$^{4, 1}$,
Henrique~Reggiani$^{5}$ \\
$^1$Research School of Astronomy and Astrophysics, Australian National University, Canberra, ACT 2611, Australia\\
$^2$ARC Centre of Excellence for All Sky Astrophysics in 3 Dimensions (ASTRO 3D), Australia\\
$^3$Department of Astronomy, Stockholm University, AlbaNova University Center, SE-106 91 Stockholm, Sweden\\
$^4$Stellar Astrophysics Centre, Department of Physics and Astronomy, Aarhus University, Ny Munkegade 120, DK-8000 Aarhus C, Denmark\\
$^5$The Observatories of the Carnegie Institution for Science,
813 Santa Barbara St., Pasadena, CA 91101, USA
}
\date{\today}
\maketitle

\begin{abstract}
The detection of $\lisi$ in Spite plateau stars contradicts the standard Big Bang nucleosynthesis prediction, known as the second cosmological lithium problem.
We measure the isotopic ratio $\ratio$ in three Spite plateau stars: HD 84937, HD 140283, and LP 815-43. We use 3D NLTE radiative transfer and for the first time apply this to high resolution, high-S/N data from the ultra-stable VLT/ESPRESSO spectrograph. These are amongst the best spectra ever taken of any metal-poor stars. 
As the measurement of $\ratio$ is degenerate with other physical stellar parameters, we employ Markov chain Monte Carlo methods to find the probability distributions of measured parameters. As a test of systematics we also use three different fitting methods. 
We do not detect $\lisi$ in any of the three stars, and find consistent results between our different methods. We estimate 2$\sigma$ upper limits to $\ratio$ of 0.7\,\%, 0.6\,\%, and 1.7\,\% respectively for HD 84937, HD 140283, and LP 815-43.
Our results indicate that there is no second cosmological lithium problem, as there is no evidence of $\lisi$ in Spite Plateau stars. 
\end{abstract}


\section{Introduction}
Lithium is the only element with three production channels: Big Bang nucleosynthesis (BBN), cosmic rays, and stars. As such, lithium is an unique element which can be used to study chemical evolution in multiple environments and processes simultaneously. Standard BBN produced mostly hydrogen and helium, with trace amounts of $\lise$, and even less $\lisi$. The predicted lithium production in standard BBN is A($\lise$)$=2.75 \pm 0.02$ and $\ratio \approx 10^{-4}$ \citep{pitrou18}. Cosmic rays and stellar production of lithium is nearly negligible in the early Universe, with BBN being the dominant source \citep{prantzos12}.

Old metal-poor dwarf stars in the Milky Way halo have been found to exhibit similar photospheric lithium abundance over a wide range of metallicities, known as the Li Spite plateau: A$(\lise)=$2.0--2.2 \citep{spite82, bonifacio97, ryan99, asplund06, melendez10, sbordone10}. While this was long thought to represent the primordial value, measurements of anisotropies of the cosmic microwave background (CMB) by the WMAP and Planck satellites yield a very precise baryon density of the Universe \citep[e.g.][]{planck2020}, which implies a significantly higher predicted Li abundance from standard BBN by a factor of 3-4 than observed in the oldest stars. This is the long-standing  cosmological lithium problem.
\citep[see e.g.,][and references therein]{fields11}. 
A possible stellar evolution resolution to the cosmological lithium problem would be that the metal-poor stars on the Spite plateau have depleted most of their $\lise$ throughout their pre-main sequence and main sequence evolution \citep{richard05, korn06, fu15}.

Not only do measured $\lise$ abundances disagree between BBN predictions and stellar observations, but so do apparently $\lisi$ abundance measurements. There are multiple claimed detections of $\lisi$ in Spite plateau stars at a level of $\ratio$ of a few percent \citep[e.g.,][]{smith93, smith98, hobbs94, cayrel99, nissen99, nissen00, asplund06, steffen12}, but at a level greatly exceeding predictions from standard BBN theory as well as production by cosmic rays or in stars \citep{prantzos12}. 
Since $\lisi$ is even more fragile than $\lise$, a depletion of $\lise$ in the surface layers of metal-poor stars to resolve the cosmological Li problem would imply an even greater depletion in $\lisi$ \citep{richard02}. As a result, stellar evolution by itself is not enough to resolve the discrepancy between BBN predictions and measurements in old halo stars for both $\lise$ and $\lisi$. This additional tension due to $\lisi$ is known as the second cosmological lithium problem. \bba{To alleviate these discrepancies between $\lisi$ measurements and predictions, various exotic scenarios have been put forward \citep[e.g.,][]{kusakabe08, jedamzik09, luo21}.}

The determination of \bba{the isotopic ratio\footnote{$\ratio = 10^{\alisi - \alise} \equiv \frac{N(\lisi)}{N(\lise)}$, where $N$ is the number density}} $\ratio$ is extremely challenging as it relies on measuring the small additional spectral line asymmetry introduced by the small isotope shift ($\approx 0.017$\,nm) in the \ion{Li}{i} 670.8\,nm resonance line when $\lisi$ is present. The 670.8\,nm line is already intrinsically asymmetric due to its unresolved doublet and fine structure components, which is further perturbed by stellar surface convection and rotation. Exceptionally high-quality observations in terms of S/N and spectral resolving power is therefore required as well as realistic modelling of the stellar atmosphere and line formation process. 
Early measurements of $\lisi$ were carried out using 1D hydrostatic simulations of stellar atmospheres, and with radiative transfer computed under the assumption of local thermodynamic equilibrium (LTE). However, convective velocity fields in 3D hydrodynamic stellar atmospheres can mimic the appearance of $\lisi$
\citep{asplund06, cayrel07}. To further complicate the picture, line formation under non-LTE (NLTE) influences the line opacity and thus the formation depth \citep{asplund03}, which also affects the measured $\lisi$.

\bba{Pioneering work on measuring $\ratio$ with 3D line formation and considering departures from LTE have been performed by \citet{cayrel08}, \citet{asplund08} and \citet{steffen10a, steffen10b, steffen12}, but unfortunately led to somewhat conflicting results. The first analysis that investigated 3D NLTE line formation for the Li line and other calibration lines consistently was performed by \citet{lind13} and resulted in non-detections of $\lisi$ at 2$\sigma$ in all investigated stars. However, they could not rule out at higher significance that the lighter isotope is present in HD 84937 at a 1.1--1.7\,\% level depending on analysis assumptions. In this study, we present new data and improved analysis techniques to further reduce the error bars and reveal if $\lisi$ is indeed present in detectable amounts in some metal-poor halo stars.}

As the isotopic splitting in lithium is small, the absorption due to $\lise$ introduces only a small asymmetry in the $\lisi$-dominated line profile. It is therefore necessary to simultaneously know the centre, depth and width of the line profile through estimates of the radial velocity, overall Li abundance, and rotational, convective and instrumental broadening, resulting in a partial degeneracy between these parameters. 
The rotational broadening and radial velocity can be measured using other lines, termed calibration lines. However, velocity fields in 3D hydrodynamic stellar atmospheres and NLTE effects affect the measured rotational broadening and radial velocity from every line. Therefore, using 3D NLTE radiative transfer for both calibration lines and the Li line is important to accurately measure $\ratio$ as shown by
\citet{lind13}. 
3D NLTE radiative transfer make detections less significant in comparison to LTE.

In the present work, we revisit the measurement of $\lisi$ in the three Spite plateau stars HD 84937, HD 140283, and LP 815-43, using improved observations, highly realistic stellar atmosphere and line formation modeling, and more sophisticated inference methods. 
We selected two bright targets from \citep{asplund06}: HD 140283 and LP 815-43, where the former subgiant 
is the brightest very metal-poor star known and hence affords exceptional-quality observations, 
and the latter is the most metal-poor star with a claimed $\lisi$ detection.
We also observe the bright benchmark turn-off star HD 84937 which has consistent previous $\lisi$ detections \citep[see e.g.][]{smith93, smith98, hobbs94, cayrel99, asplund06}; in particular, HD 84937 has been studied in 3D NLTE by \citet{steffen12} who detected $\lisi$ and \citet{lind13} who did not detect $\lisi$.
Observations of these stars were made using ESPRESSO/VLT \citep{pepe10} which was designed for extremely high spectral fidelity, in particular in terms of the stability of its wavelength solution.
We accurately model the synthetic spectra through the use of 3D hydrodynamic model atmospheres and NLTE radiative transfer for all lines used in this work, both Li and calibration lines of K and Fe. Only recently have 3D NLTE profiles for such more complex elements become feasible 
\citep{amarsi16b}. In addition, we use a Markov chain Monte Carlo method to sample the posterior distribution of $\ratio$, \bba{which} consistently takes into account the complicated partial degeneracies between parameters.

In this paper, we show the observations in Section~\ref{sec:obs}, describe our methodology in Section~\ref{sec:method}, present our results in Section~\ref{sec:res}, and finally discuss these results in Section~\ref{sec:diss}.

\section{Observations}
\label{sec:obs}
Observations were taken with the ESPRESSO \citep{pepe10,pepe20} high resolution spectrograph on the ESO Very Large Telescope (VLT) in the high resolution (HR) 1-UT configuration with 1x1 binning. Observations were executed during April to July 2019 in service mode, with good seeing (typically less than $0.7\arcsec$), and data were reduced using the ESPRESSO pipeline version 1.3.2. 
Exposure times were 7$\times$50\,minutes for HD 84937 and 24$\times$50\,minutes for LP 815-43. 
Exposures were shorter, $4 \times 20$\,minutes for our brightest target, HD 140283, to avoid saturation.
The spectra are of exceptional quality, with a stacked signal-to-noise ratio $S/N > 1000$ per pixel and a resolving power of $R \approx 146\,000$ in the region of the lithium line. 

For each star, we align exposures according to the radial velocity reported by the pipeline. 
We perform the coaddition using a flux-weighted summation with $3\sigma$ outlier rejection, on a merged uniform wavelength grid with a pixel step of 0.7\,\kms, corresponding to Nyquist sampling at the nominal resolution of the spectrograph.
We note that this procedure will partly correlate neighbouring pixels and hence standard $\chi^2$ statistics is formally not appropriate.

The photon-limited radial velocity precision for stellar spectra that are rich in sharp spectral lines approaches 0.1\,\ms\, thanks to the combined use of a ThAr lamp and Fabry-Per\'ot interferometer that cover the entire optical wavelength range, and the temperature and pressure stabilisation of the instrument \citep{pepe20}. We stress however that this internal precision is different from the accuracy of the instrument, which is better characterised by the accuracy of the wavelength solution at the level of 20\,\ms\ with significant dependence on wavelength \citep{schmidt_fundamental_2020}. 

We use the ThAr calibration frames associated with our observations to estimate the resolving power of the instrument, and fit the variation of the instrumental FWHM as a function of pixel number using 2nd-order polynomials. We note that as the ESPRESSO instrument uses an amorphic pupil slicing unit, each physical diffraction order is recorded twice with slightly different optical characteristics. 
Due to the limited number of ThAr calibration lines in each spectral order, we smooth our polynomial fits to the resolving power across adjacent diffraction orders. The resulting smoothed mapping agrees with individual ThAr measurements at the level of 0.1 and 0.3\,\% on the blue and red detectors, respectively. 

Like \citet{pepe20}, we find systematic differences between the blue and red cameras, as well as between the extracted slices for a given diffraction order, and a tendency for higher resolving power toward higher wavelengths both across each spectral order and comparing spectral orders across each camera. 
For example, we estimate $R=146400$ near \liline, $R=153000$ near \kline, and $R=137000$, 142000 and 142000 respectively at 540, 511 and 490\,nm where there are numerous \feI\ lines.
We produce a single estimate of the resolving power for each extracted pixel in our merged, coadded spectra using the relative throughput of each extracted slice as estimated using the flat field calibration.

We emphasise that our observed stellar spectra have higher resolving power, spectral sampling, and $S/N$ than any previously obtained for metal-poor stars.

\subsection{Residual interference pattern}
\label{sec:interference}
Due to the exceptional quality of our spectra, the limited quality of calibration flat field frames leaves residual patterns in the reduced data. In particular, an interference pattern has an apparent amplitude greater than the Poisson noise in the co-added spectra. Thanks to the excellent stability of the instrument, we found no significant variation in flat fields during a given month and therefore created monthly master flat frames using the full set of flat field exposures associated with our observations. Despite this, a residual pattern with a period of roughly 0.1\,nm and an amplitude of $\approx 0.1$\% is apparent in our data, similar to what has been found in the literature \citep{allart_wasp-127b_2020,casasayas-barris_atmosphere_2021}.

We found limited success with applying Fourier transforms and explicitly fitting periodic functions to the data, but were able to identify and remove the interference pattern using Gaussian processes.
Gaussian processes have been applied to many areas in astronomy, usually to fit complex data \citep[e.g.,][]{aigrain_k2sc_2016, angus18, iyer19, hu20, feeny21, soo21}, 
but rarely to correct interference patterns in spectra.
A full description of Gaussian processes is beyond the scope of this work, for the interested reader, we refer to \citet{rasmussen2006gaussian}.

We test Gaussian processes on several nearly line-free regions, with good results when extracting a spectral region of 1200\,\kms\ (roughly 1700 pixels), with any absorption lines masked out. 
We fitted the oscillation pattern with Gaussian processes using the Python package \sklearn\ \citep{scikit-learn}. For this fit, we normalised the wavelength scale, shifted the spectrum to 0 and smoothed it using a rolling mean with a width of 11 pixels. We also experimented with other widths for the rolling mean with minimal difference between 5, 11, and 21 pixels. 
We used a periodic kernel multiplied by a radial basis function kernel. The periodic kernel is given by
\begin{equation}
k(x_i, x_j) = \exp{\left(-\frac{2\sin^2(\pi d(x_i, x_j)/p)}{l^2}\right)},
\end{equation}
where $l$ is the length scale which determines how correlated neighbouring points are, $p$ is the period which determines the periodicity of the pattern, and d is the Euclidean distance which in this work simplifies to $d(x_i, x_j) = \vert x_i - x_j \vert$ as $x_i$ and $x_j$ are scalars; the radial basis function kernel is given by
\begin{equation}
k(x_i, x_j) = \exp{\left(-\frac{d(x_i, x_j)^2}{2l^2}\right)}.
\end{equation}
We found a reasonable fit with $l = p = 1$, representing a relatively short periodicity for the pseudo-periodic variation, and multiplied this periodic kernel by a radial basis function kernel with $l = 500$ that absorbs minor ripples in the continuum. 

Due to the flexibility of Gaussian processes, the solution tends to oscillate in masked regions that were not included in the fit. We therefore verified the robustness of the method on a line-free region near 680\,nm, where we applied an identical line masking as was used on the lithium-line region at 671\,nm, and found that we were able to predict the interference pattern in the masked out section. This is discussed further in App.~\ref{sec:GP_test}. On the contrary, we found that we could not robustly fit the residual interference pattern in the region of the potassium line at 770\,nm due to the presence of telluric lines, nor at shorter wavelengths due to the presence of many weak lines with relative depths in excess of $10^{-3}$.

\begin{figure*}
    \centering
    \includegraphics[width=\textwidth]{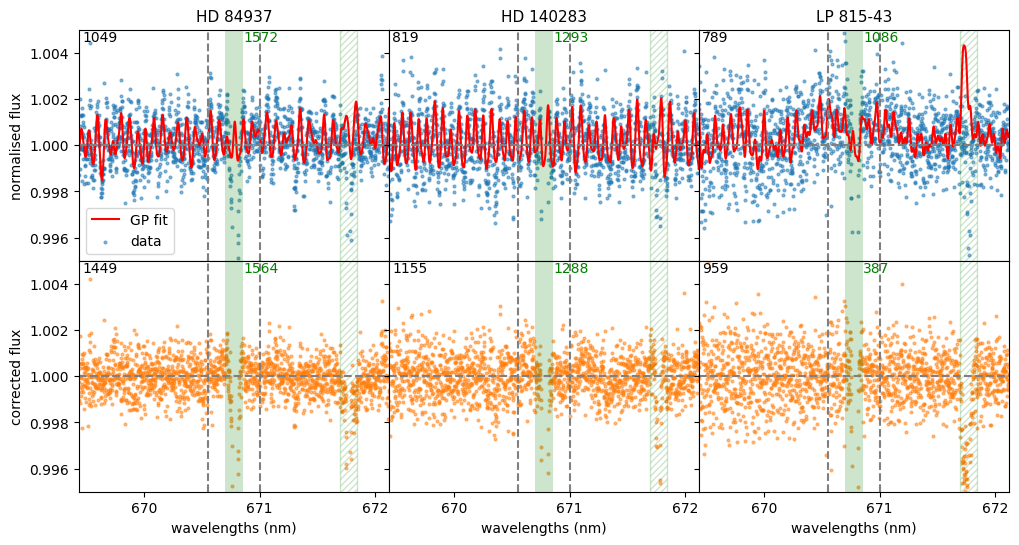}
    \caption{The residual interference pattern in the \liline\ line in HD 84937 (left), HD 140283 (middle), and LP 815 (right). Blue shows the stacked observation with the residual interference pattern, red shows the Gaussian process fit to this residual interference pattern, and orange is the stacked spectra with the residual interference pattern removed. Green shows the region which was not considered in the fit. The S/N for the stacked continuum given by the MAD of the non-masked region is shown in black on the top \bba{left}. The S/N for the Li line (green shaded region excluding hatched region) given by median of the flux divided by flux error is shown in green on the top middle. The vertical grey dashed lines show the region used for continuum normalisation.}
    \label{fig:interference}
\end{figure*}

We show in Fig.~\ref{fig:interference} the lithium line region at 671\,nm for all three stars, together with our fit to the interference pattern. 
While the interference pattern is relatively periodic and nearly sinusoidal in the spectrum of HD 140283, the spectrum of LP 815-43 appears nearly chaotic. We note that while spectra for HD 140283 were recorded over the span of two hours, observations for LP 815-43 span seven weeks and likely exhibit some time variability in the coadded spectrum.
Due to the interference pattern, the pixel-to-pixel scatter implies a S/N that is significantly worse than what is estimated from Poisson statistics.
After subtracting the fitted oscillation pattern, the corrected spectra do not appear to be systematically distorted and exhibit a significantly boosted S/N in the continuum. 
As Gaussian processes provide an error estimate associated with the model prediction, we fold these error estimates into our final S/N estimates. The error in the Li line region for LP 815-43 after folding in the Gaussian process error is significantly \bba{larger} than the original S/N. Therefore, we use the normalised flux not the corrected flux for our analysis of LP 815-43.

\section{Methodology}
\label{sec:method}
\subsection{3D hydrodynamic model atmospheres}
The fundamental stellar parameters of all three stars are accurately known from advanced spectroscopic or fundamental measurements.
The effective temperature ($\teff$) of HD 140283 is accurately known from fundamental measurements of its angular diameter and bolometric flux \citep{karovicova20}; for HD 84937 and LP 815-43 we adopt Balmer line fits based on 3D NLTE spectrum synthesis \citep[from][]{amarsi19}. 
Additionally, the surface gravity $\logg$ is well constrained for all three stars by parallax measurements from Gaia DR2 \citep{gaiadr2} or the Hubble Space Telescope \citep{vandenberg14}, together with masses from stellar evolution models. 
The metallicities of HD 84937 ($\rm [Fe/H] = -2.05 \pm 0.06$) and LP 815-43 ($\rm [Fe/H] = -2.68 \pm 0.06$) were determined using 3D LTE spectrum fits to unblended \feII\ lines by \citet{amarsi19}.
For HD 140283, we use the value $\rm [Fe/H] = -2.29 \pm 0.11$ from \citet{karovicova20} that is based on 1D NLTE modeling of a set of unblended \feI\ and \feII\ lines; we note that this value is in good agreement with the measurement based on 3D LTE modelling of \feII\ lines by \citet{amarsi19}, $\rm [Fe/H] = -2.36 \pm 0.05$, even though the latter used slightly different values for $\teff$ and $\logg$.

We use 3D hydrodynamic atmospheric simulations computed with the STAGGER code, 
and list the stellar parameters of our models and the targeted stars in Table~\ref{tab:sparams}.
Our models were computed on a staggered Cartesian mesh with $240^3$ volume elements, using a horizontally periodic boundary box; the top and bottom boundaries are open, with the entropy and thermal pressure of incoming matter at the bottom tuned to produce the desired $\teff$. 
\bba{We computed models of} HD 84937 and HD 140283 specifically for this project, while for LP 815-43 we used a model from the Stagger-grid \citep{magic13}.
The tailored models were computed consistently with \citet{magic13}, with only minor updates to the code as described by \citet{collet18}; these changes do not crucially change the outcome of the modelling. All models use the metal mixture from \citet{asplund09}.

\begin{table*}
    \centering
    \caption{Stellar parameters for the programme stars from literature, and the corresponding properties of the 3D model atmospheres; for the latter, the error in $\teff$ represents the standard deviation of the $\teff$ time series, as effective temperature fluctuates over time in 3D hydrodynamic model atmosphere.}
    \begin{threeparttable}
    \begin{tabular}{l|c|c|c}
    & HD 84937$^\textrm a$ & HD 140283$^\textrm b$ & LP 815-43$^\textrm a$ \\
    \hline 
    $\teff$ (K) literature & $6340 \pm 70$ & $5792 \pm 55$ & $6461 \pm 70$  \\
    $\teff$ (K) model & $6371 \pm 20$ & $5774 \pm 11$ & $6504 \pm 20$ \\
    \hline
    $\logg$ literature & $4.05 \pm 0.06$ & $3.65 \pm 0.02$ & $4.11 \pm 0.06$ \\
    $\logg$ model & 4.00 & 3.70 & 4.00 \\
    \hline
    $\met$ literature & $-2.05 \pm 0.06$ & $-2.29 \pm 0.11$ & -2.68 $\pm 0.06$  \\
    $\met$ model & -2.00 & -2.50 & -3.00
    \end{tabular}
    \begin{tablenotes}
    \item [a] \citet{nissen14, amarsi19} 
    \item [b] \citet{karovicova20}
    \end{tablenotes}
    \end{threeparttable}
    \label{tab:sparams}
\end{table*}

\subsection{3D NLTE radiative transfer}
We performed NLTE radiative transfer calculations using the \balder\ code \citep{amarsi16a, amarsi16b, amarsi18}, which originates in the \multitd\ code \citep{botnen99,leenaarts09}. 
The model atmospheres were interpolated to a smaller number of volume elements, $120^2$ by 220 (vertical), for computational efficiency reasons.
We selected at least five snapshots from each hydrodynamic simulation, chosen equidistantly to sample at least one convective turnover time. 
This selection results in an error of roughly 0.01\,dex in the determination of $\ali$. The variations in inferred radial velocity due to convective motions in the 3D stellar models are somewhat harder to estimate, and appears to differ significantly between models; in no case does this induce an error ($\sigma / \sqrt N_\text{snapshots}$) greater than 150\,\ms -- in the best case in fact just 7\,\ms\bba{. We} note \bba{that} the accuracy in radial velocity is significantly better than this, as different spectral lines exhibit a similar radial velocity for a given point in time in the hydrodynamic simulation. 
For the NLTE radiative transfer solution, we use 26 rays for Li, representing two vertical rays and six inclined angles rotated across four azimuthal directions; for Fe and K lines used for parameter determination (see Section~\ref{sec:spec_fit}) we use only two inclined angles, for a total of 10 rays.
After the NLTE solution converged, we computed the emergent flux for Li with one vertical and seven inclined and eight azimuthal angles, for a total of 57 rays; for Fe and K we used only four azimuthal angles, for a total of 29 rays.

We compute synthetic profiles for three elements: Li, K, and Fe. The Li atom is from \citet{wang_3d_2020}, but with line components for $\lisi$ in addition to $\lise$; wavelengths for both isotopes were taken from \citet{smith98}, and we note that these are in excellent agreement with later ultra-precise measurements by \citet{sansonetti11}. 
We adopt the K atom from \citet{reggiani19}, whose line wavelengths originate with \citet{falke06}. 
The Fe atom is from \citet{amarsi16b}, with wavelengths originating from \citet{nave94}; we systematically correct \bba{the Fe} wavenumbers by $(6.7 \pm 0.8) \times 10^{-8}$  according to the improved measurements of reference wavelengths for \ion{Ar}{ii} lines that was identified by \citet{whaling95}.

For both \liI\ and \kI, the wavelengths of the resonance lines are known to incredible accuracy, 0.015 and 0.06\,\ms, respectively.
In comparison, the wavelengths of \feI\ lines are relatively uncertain with an accuracy of the order 25\,\ms.

\subsection{Spectrum fitting}
\label{sec:spec_fit}
Measuring the isotopic ratio from the \liline\ transition between the ground state and first excited level requires knowledge of the total Li abundance, as well as global parameters representing rotational broadening, and radial velocity;
we note that the intrinsic thermal, pressure and convective line broadening are directly computed from the 3D stellar atmosphere models (no ad-hoc macro- or microturbulence parameters necessary in any 1D modelling enter the 3D calculations) while the instrumental broadening is measured from the ThAr wavelength calibration spectra. 
We restrict $\vsini \geq 0$, but similar to other works \citep{smith93, asplund06, hernandez19} do not set any constraints on $\ratio$ as negative ratios allow us to capture potential systematic errors in modelling and avoids artificially skewing the results to positive $\ratio$ values.
Although unphysical, negative $\ratio$ values have been implemented with negative opacity contribution from the $\lisi$ isotopic component. 
These constraints are used for all fitting methods. 
In addition, prior to spectrum fitting we normalise the continuum using surrounding pixels not affected by the Li line 
in contrast to other works which fit the continuum simultaneously with other free parameters. \bba{Fitting for the continuum simultaneously has a negligible affect on our measured $\ratio$ of order 0.01\,\% due to the high S/N of the observed spectra.} \bba{For every line used in this work, we} normalise the continuum by selecting a 200--400\,\kms\ region around the line,
masking out all lines in the region, and dividing out a Theil-Sen slope fit.

Both rotational broadening and instrumental broadening are applied to the synthetic profiles. We emphasise that so-called macroturbulent broadening is not a free parameter because it arises naturally from large-scale velocity fields in the 3D hydrodynamic models \citep[e.g.][]{asplund00, nordlund09}.
To broaden our profiles, we first apply rotational broadening to the synthetic profile, following the method described in \citet{dravins90}, then we apply instrumental broadening as a Gaussian with FWHM estimated from the measurements of calibration ThAr lines, that varies with wavelength as described in Sect.~\ref{sec:obs}. 

\bba{The ESPRESSO data reduction pipeline estimates radial velocities based on cross-correlations with a template. We have derived corrections that place the spectra at rest with respect to our 3D NLTE spectra. Hence, we label the measured 3D NLTE radial velocity as $\tvr$, and the difference between our 3D NLTE radial velocity and the pipeline reported radial velocity as $\vr$.}

\subsubsection{Method 1: Li-only}
As the four parameters which contribute to the shape of the Li line are partially degenerate, it is necessary to sample the posterior distribution from a $\chi^2$ likelihood to determine the values for the parameters. We use MCMC from the Python package \texttt{emcee} \citep{foreman-mackey13} for sampling in this work. We map the sample onto 2D error ellipses using the Pearson correlation coefficient to calculate the eigenvalues which determine the radii of the ellipse, then scale the ellipse such that it encloses 1, 2, and 3$\sigma$ of the data (see Fig.~\ref{fig:ellipse}).

In addition to this method where all four parameters are determined directly from the \liline\ line, we employ two additional methods where we determine rotational broadening and radial velocity externally through other spectral lines.

\subsubsection{Method 2: K-constrained}
\label{sec:K_cons}
In order to minimise the effects of parameter degeneracies, we utilise the \kline\ line as a calibration line to determine the rotational broadening and radial velocity prior to fitting the isotopic abundances to the Li line. 
This is because it has similar properties to the \liline\ line in terms of being the resonance line of an alkali of  comparable strength in our metal-poor stars. Furthermore, the doublet structure of the \ion{K}{i} resonance lines is fully resolved (the other doublet line being at 766.5\,nm, which unfortunately is destroyed by telluric lines) in contrast to the Li doublet, which makes the line profile easier to model. As the \kline\ line also has an accurately known wavelength and exhibits only negligible isotopic and hyperfine splitting significantly less than 1\,\kms\ \citep{falke06}, it acts as an excellent reference line.

The K-constrained fit is performed by first measuring $\vr$ based on the shift of the center of the \kline\ line, this is done by fitting a Gaussian to both the observed and synthetic profile; $\vsini$ is fitted based on this previously measured $\vr$. 
The probability distributions of $\vsini$ and $\vr$ were determined by exploring a range for each parameter: for each value tested, we determine the optimal $\chi^2$ value found through optimising the other parameters, e.g: in the case of $\vsini$, we fix $\vsini$ and optimise $\vr$ and the abundance of K to find the optimum $\chi^2$. 
We prescribe values for $\vr$ and $\vsini$ from these probability distributions in a Monte Carlo, then use \texttt{emcee} to derive A(Li) and $\ratio$ from the \liline\ \bba{line}. Note that $\vsini$ and $\vr$ are dictated from calibration lines externally rather than implemented as priors to the MCMC fit on Li -- we do this to prevent the Li line from contributing to the determination of $\vsini$ and $\vr$.

\begin{figure*}
    \centering
    \includegraphics[width=\textwidth]{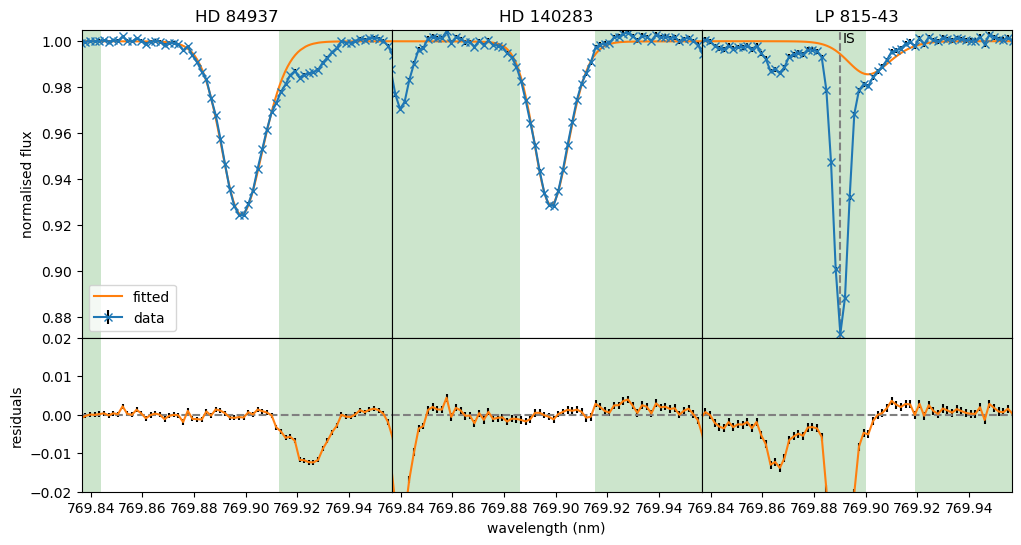}
    \caption{Observed (blue) and fitted (orange) line profile for the \kline\ line for the three observed stars HD 84937 (left), HD 140283 (middle), and LP 815-43 (right). The data points within the shaded region are not considered in the fit. The interstellar line in the LP 815-43 spectra is labelled as IS, all other lines are tellurics.}
    \label{fig:k_fits}
\end{figure*}

The best fit to K is shown in Fig.~\ref{fig:k_fits}. The residuals for the fits to the K line for HD 84937 and HD 140283 are small for the pixels used in fitting. Using our best fit and \bba{the solar abundance $\ak_\odot = 5.07 \pm 0.03$ } from \citet{asplund21}, we derive \bba{$\rm [K/H] = -1.76 \pm 0.03$ and $-2.27 \pm 0.03$ respectively for HD 84937 and HD 140283. 
This gives $\rm [K/Fe] = 0.29 \pm 0.07$ and $0.02 \pm 0.11$} for HD 84937 and HD 140283 respectively using the literature $\rm [Fe/H]$ values from Table.~\ref{tab:sparams}. \bba{We note that these abundances are higher by 0.16 and 0.03\,dex, respectively, compared to the 1D non-LTE analysis by \citet{reggiani19}. As the Galactic chemical evolution of K is not the goal of this analysis, we do not delve further into the implications of these abundance measurements. } 

We do not make a K constrained fit to LP815-43 as it has a small radial velocity, resulting in a blend between the stellar and interstellar \kline\ lines. Masking out the interstellar line removes the minimum of the stellar K line, which results in a poorly constrained fit. 
We tried fitting a Gaussian and synthetic profile to the interstellar and stellar K lines combined, however this also failed due to the telluric line blueward of the stellar \kI\ line which leaves no nearby continuum on the blue side to fit.

\subsubsection{Method 3: Fe-constrained}
Our spectra contain numerous unblended \feI\ lines which offer a statistical estimation of $\vsini$ and $\vr$. The method employed here is similar to the K-constrained method discussed in the previous section, differing in that we measure 47 \feI\ lines compared to only one \kI\ line.

We select a set of unblended \feI\ lines whose strengths are comparable to the \liline\ line. We estimate their distributions 
using the same method as the K line. Since we have 47 Fe lines, we measure 47 distributions for $\vsini$ and $\vr$ respectively. To combine these 47 distributions, first, the median of each distribution is used to remove outlier distributions. Next, the distributions are each scaled with respect to their $\chi^2_{red}$ value and finally combined together into a joint probability distribution by summing the distributions, known as a mixture model.

\begin{figure*}
    \centering
    \includegraphics[width=0.7\textwidth]{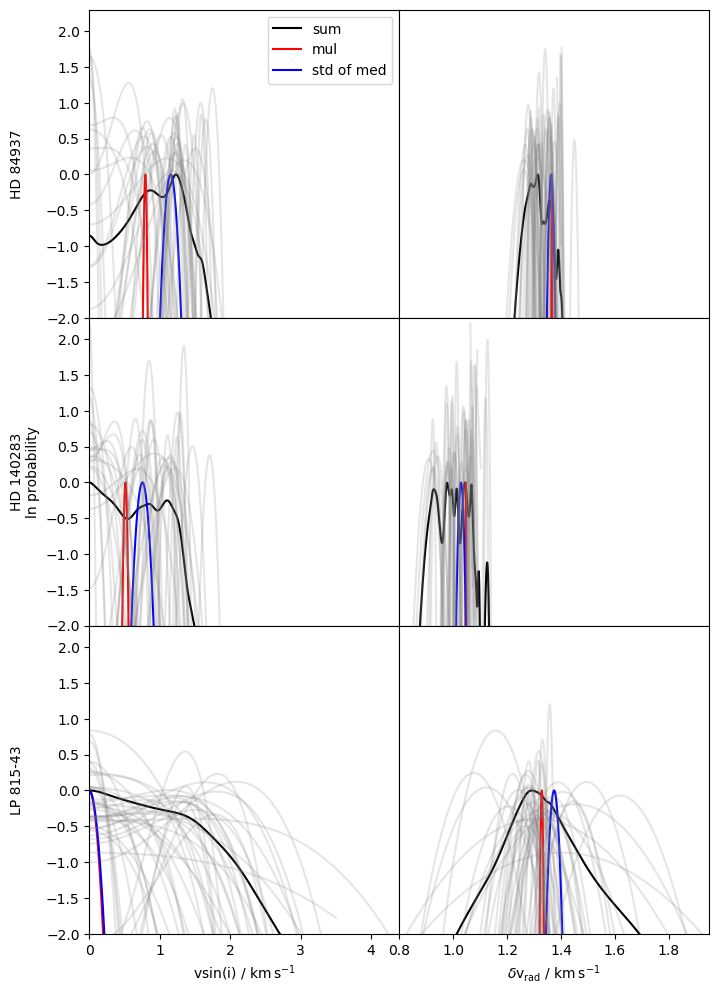}
    \caption{The log probability distributions without outliers of $\vsini$ (left) and $\vr$ (right) from \feI\ lines for HD 84937 (top), HD 140283 (middle), and LP 815-43 (bottom). The individual probability distributions (grey) shown are scaled with respect to their $\chi^2_{red}$ value. The different joint probability distributions are also shown: mixture model, which is the summation of the individual probability distributions (black), multiplication of the individual probability distributions (red), and the standard method of using the median and standard error \bba{of the optimums} as the mean and standard deviation of a Gaussian (blue).}
    \label{fig:fe_dist}
\end{figure*}

Fig.~\ref{fig:fe_dist} shows the probability distribution of $\vsini$ and $\vr$ measured from each Fe line and the joint probability distributions computed using three different methods. The standard method to find the joint probability distribution for $\vsini$ and $\vr$ given some lines would be to estimate the optimum parameters from the lines, then compute the mean or medium and take the standard error of the \bba{optimums} as the error estimate (shown in blue in the figure). However, $\vsini$ is only marginally resolved in slowly rotating stars, therefore, the asymmetry of the distributions of $\vsini$ need to be taken into account. Considering the asymmetry, we compute the probability distribution of each parameter for each line. To get the joint probability distribution, the usual method is to multiply each individual probability distribution together (shown in red in the figure). In the case where individual probability distributions have similar error, and this error is larger than the scatter the in optimum, then the mean and error of the individual probability distributions multiplied together is equivalent to the mean and error given by the standard method. In this work, we find that the scatter in $\chi^2$ optimum is much larger than the error in individual distributions, as shown by the individual probability distributions (shown in grey in the figure). Therefore, we choose to use a mixture model as the joint distribution (shown in black in the figure). The physical interpretation of this mixture model is that each Fe line measures $\vsini$ and $\vr$ with some probability. This joint distribution is used to measure $\ali$ and $\ratio$ from the \liline\ line using the same method described in Section~\ref{sec:K_cons}.

\begin{figure*}
    \centering
    \includegraphics[width=\textwidth]{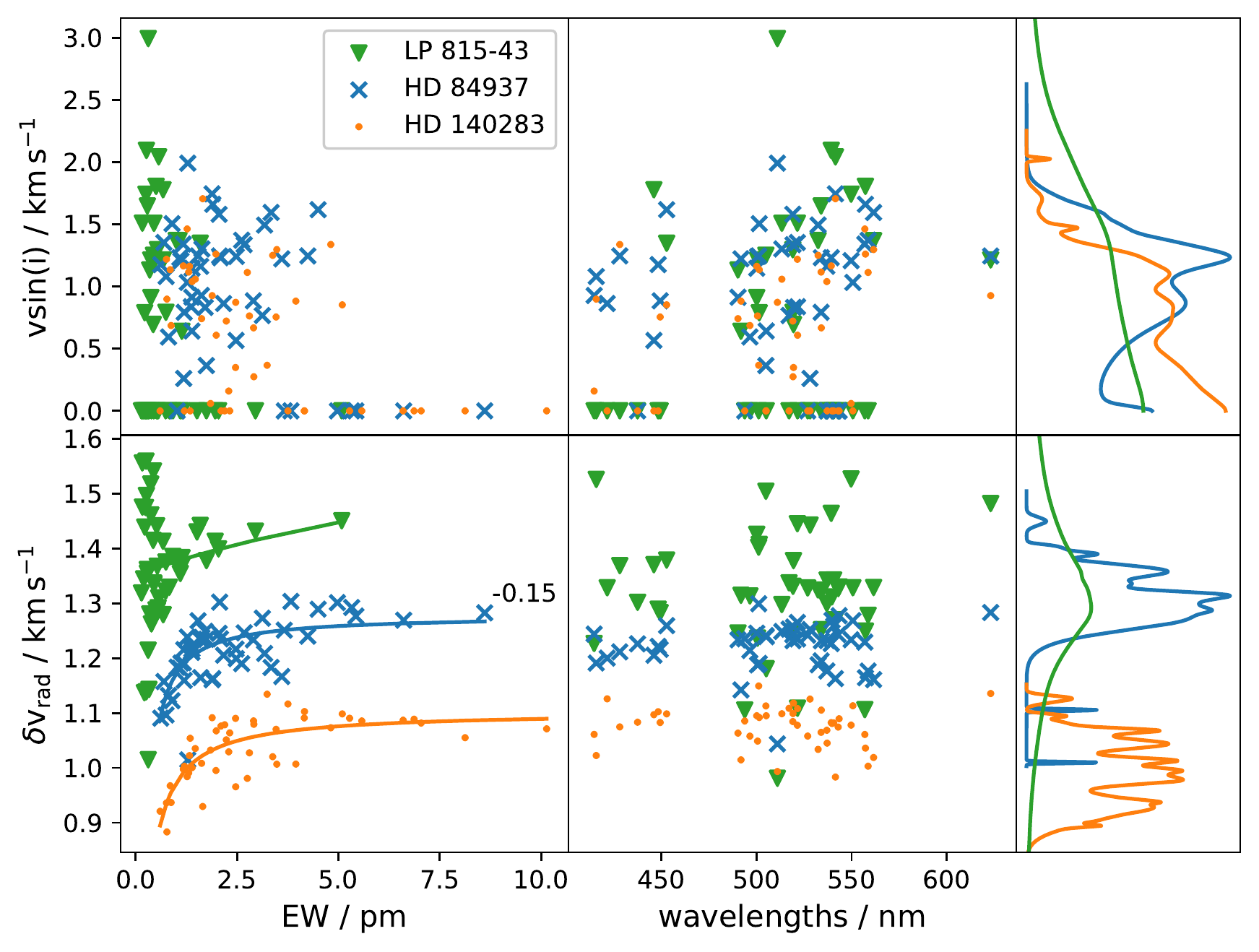}
    \caption{Comparison of $\vsini$ and $\vr$ against EW and wavelength for fits to a set of \feI\ lines. 
    We note that $\vr$ was shifted \bba{by -0.15} for HD 84937 in order to avoid overlap with LP 815-43. 
    In the comparison between $\vr$ and EW, a fit to the data illustrates that the measurements correlate. In the comparison between $\vr$ and wavelength, this correlation has been removed for illustration purposes, and the data have been shifted by arbitrary amounts in order to avoid overlap.
    The rightmost panels show the normalised distributions of $\vsini$ and $\vr$. 
    Error bars have been omitted from this figure in order to avoid crowding, but can be seen in Fig.~\ref{fig:fe_dist}. 
    }
    \label{fig:fe_trends}
\end{figure*}

Fig.~\ref{fig:fe_trends} shows the maximum likelihood estimates of $\vsini$ and $\vr$ as a function of EW, and wavelength. 
There is a positive correlation between our estimated $\vr$ and EW. We emphasise that since our 3D NLTE modelling predict varying convective shifts for lines of different strengths, the trend seen in Fig.~\ref{fig:fe_trends} is not the well-known behaviour of decreasing convective blue-shift for stronger lines present in late-type stars like the Sun \citep[e.g.][]{allendeprieto98,allendeprieto02}
but a shortcoming in the predicted line shifts in these metal-poor stars for reasons that remain unclear.  
Removing this trend only for the wavelength to $\vr$ panel (bottom middle) for illustrative purposes shows that there is no residual trend associated with wavelength. 
We note that the zero-point of this trend is ambiguous, and removing it therefore would not improve the accuracy of our $\vr$ measurement. \bba{As a result, we use the measured $\vr$ probability distributions in our analysis without removing the trend.}
A small number of extreme points can be seen as narrow peaks in the distributions; these are not ignored, but are taken into account in our joint probability distribution.

We provide the optimum fits from the \feI\ lines and show some examples of these line fits in App.~\ref{sec:fe_fits}.

\section{Results}
\label{sec:res}
We have estimated $\ratio$ in HD 84937, HD 140283 and LP 815-43, using three different methods that involve different treatment of the associated parameters $\ali$, $\vsini$ and $\tvr$. 
We have measured all four parameters directly using the \liline\ line (Li-only), calibrated $\vsini$ and $\tvr$ using the \kline\ line (K-constrained), or likewise but using 47 \feI\ lines (Fe-constrained).
Due to an unfortunate blend between the stellar and interstellar components of \kline, we could not utilise this line in the spectrum of LP 815-43.

\begin{figure*}
    \centering
    \includegraphics[width=\textwidth]{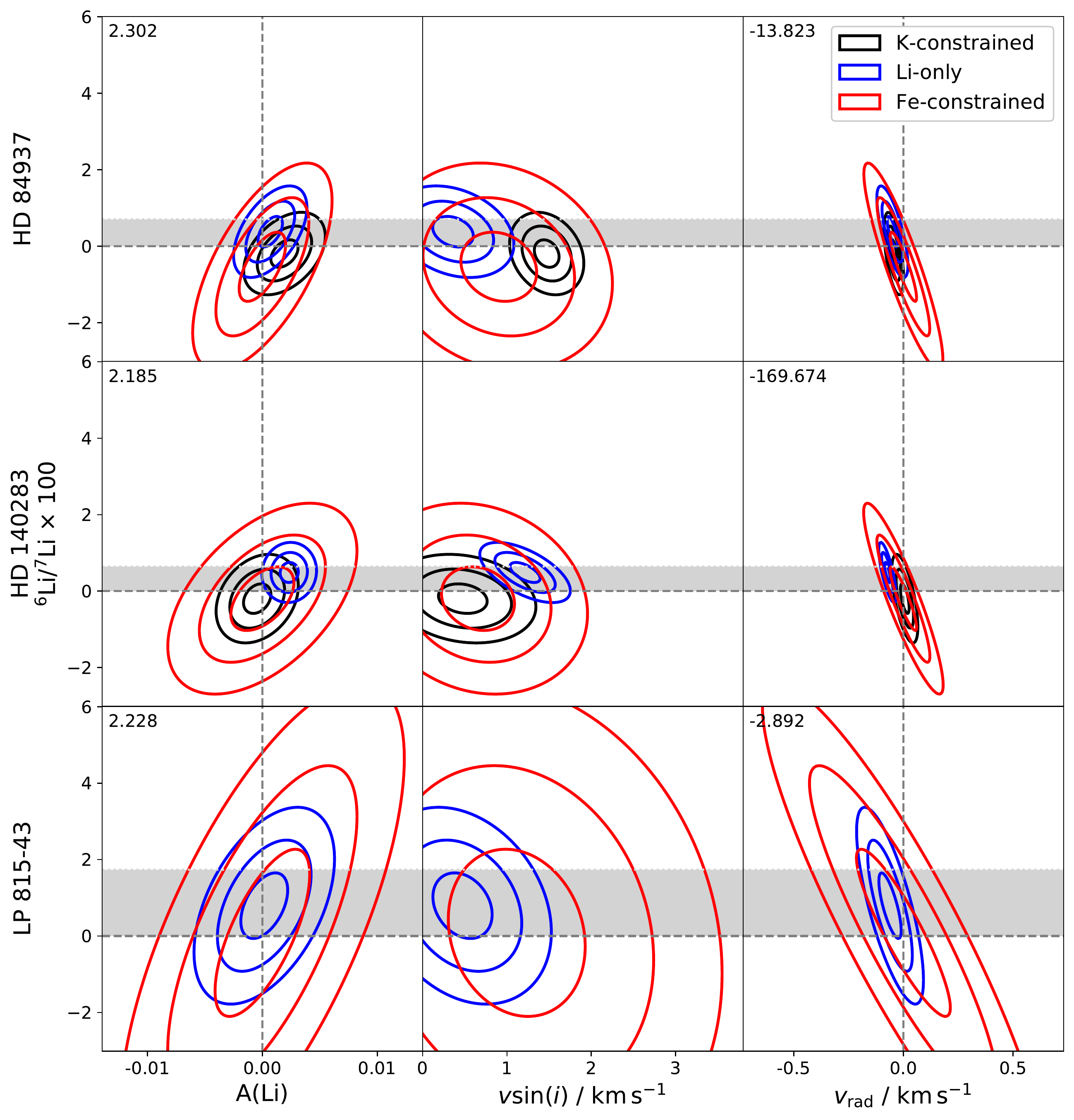}
    \caption{Confidence ellipses containing 1, 2, and 3$\sigma$ of the data. Blue ellipses shows fitting all parameters on the Li line (Li-only), black ellipses show the fit using the K line as a calibration line (K-constrained), and red ellipses show the fit using Fe calibration lines (Fe-constrained). The $\ali$ and $\tvr$ ellipses are \bba{shifted arbitrarily so they are centered close to 0}, the amount \bba{shifted} by for each panel is shown on the top left corner \bba{(same units as axes)}. K-constrained fits were not made for LP 815-43. \bba{The shaded grey region indicates our measured 2$\sigma$ upper limits, as explained in the text.}}
    \label{fig:ellipse}
\end{figure*}

We show our results in Table~\ref{tab:results} and Fig.~\ref{fig:ellipse}. In our Li-only and K-constrained fits, we find typical errors of order 0.001\,dex for $\ali$, 0.3\,\% for $\ratio$, 0.2\,\kms\ for $\vsini$, and 0.02\,\kms\ for $\tvr$; results for LP 815-43 have roughly twice as large errors as the other stars due to the exceptional quality of the spectra for the latter. We note also that the spectrum of LP 815-43 did not have sufficient quality to remove the residual interference pattern discussed in Sect.~\ref{sec:interference}. 
All three methods for all three stars are consistent with $\ratio = 0.00\%$ within 2$\sigma$: we do not detect $\lisi$ in our stars. \bba{We place upper limits using the distance between the median and 95th percentile of the $\ratio$ posterior distribution. This results in 2$\sigma$ upper limits of 0.7\,\%, 0.6\,\%, 1.7\,\% on $\ratio$ for HD 84937, HD 140283, and LP 815-43 respectively. For each star we use the most precise method, i.e., K-constrained for HD 84937 and HD 140283, and Li-only for LP 815-43.}

Constraining $\vsini$ and $\tvr$ with calibration lines \bba{moves} the median to lower isotopic ratios. Differences are however comparable to the estimated error bars, indicating that systematic errors related to the determination of nuisance parameters do not significantly influence our non-detection.
$\ali$, $\vsini$, and $\tvr$ are all partially degenerate with $\ratio$ as seen from the tilt of the ellipses. In particular, if $\ali$ increases, the $\ratio$ will also increase; 
if $\vsini$ increases, $\ratio$ will decrease; and if $\tvr$ increases, $\ratio$ will also decrease. This correlation is seen for all fitting methods, with the strength of the correlation varying depending on the method, but the direction of correlation remains the same. Notably, we do not find a significant detection of $\lisi$ in any spectrum, regardless of method. 

\begin{table*}
    \centering
    \caption{Measured $\ali$ (dex), $\ratio$ (\%), $\vsini$ (\kms), $\tvr$ (\kms), and their corresponding $\chi^2_{red}$ using different fitting methods for HD 84937, HD 140283, and LP 815-43. The reported numbers are the median, 16th percentile, and 84th percentile.}
\begin{tabular}{cc|ccc}
& & Li-only  & K-constrained & Fe-constrained \\
\hline
\hline
\multirow{5}{5em}{HD 84937} & $\ali$ & $2.303 \ -{0.001}/+{0.001}$ & $2.304 \ -{0.001}/+{0.001}$ & $2.302 \ -{0.002}/+{0.002}$ \\
& $\ratio$ & $0.4 \ -{0.4}/+{0.3}$ & $-0.2 \ -{0.4}/+{0.3}$ & $-0.6 \ -{0.8}/+{0.7}$ \\
& $\vsini$ & $0.32 \ -{0.21}/+{0.30}$ & $1.48 \ -{0.15}/+{0.13}$ & $0.94 \ -{0.57}/+{0.42}$ \\
& $\tvr$ & $-13.872 \ -{0.024}/+{0.023}$ & $-13.864 \ -{0.018}/+{0.013}$ & $-13.820 \ -{0.045}/+{0.054}$ \\
& $\chi^2_{red}$ & $1.20$ & $1.27$ & $1.23$ \\
\hline
\multirow{5}{5em}{HD 140283} & $\ali$ & $2.187 \ -{0.001}/+{0.001}$ & $2.184 \ -{0.001}/+{0.001}$ & $2.185 \ -{0.002}/+{0.002}$ \\
& $\ratio$ & $0.5 \ -{0.3}/+{0.2}$ & $-0.2 \ -{0.3}/+{0.3}$ & $-0.2 \ -{0.8}/+{0.9}$ \\
& $\vsini$ & $1.23 \ -{0.17}/+{0.16}$ & $0.46 \ -{0.31}/+{0.34}$ & $0.63 \ -{0.47}/+{0.51}$ \\
& $\tvr$ & $-169.747 \ -{0.013}/+{0.013}$ & $-169.670 \ -{0.018}/+{0.022}$ & $-169.674 \ -{0.067}/+{0.067}$ \\
& $\chi^2_{red}$ & $1.20$ & $1.47$ & $1.43$ \\
\hline
\multirow{5}{5em}{LP 815-43} & $\ali$ & $2.228 \ -{0.002}/+{0.002}$ & & $2.228 \ -{0.004}/+{0.004}$ \\
& $\ratio$ & $0.8 \ -{0.9}/+{0.9}$ & & $0.0 \ -{2.5}/+{2.2}$ \\
& $\vsini$ & $0.39 \ -{0.26}/+{0.45}$ & & $0.99 \ -{0.71}/+{0.93}$ \\
& $\tvr$ & $-2.953 \ -{0.052}/+{0.050}$ & & $-2.915 \ -{0.147}/+{0.210}$ \\
& $\chi^2_{red}$ & $1.83$ & & $1.84$ \\
\hline
    \end{tabular}
    \label{tab:results}
\end{table*}

The Fe-constrained results exhibit significantly larger uncertainties on $\vsini$ and $\tvr$, as compared to the Li-only and K-constrained results. 
As shown in Fig.~\ref{fig:fe_trends}, our measurements of $\vsini$ from \feI\ lines exhibit strong asymmetry, and are not clearly distinguished from zero. In part, this is due to a number of \feI\ lines that appear more narrow than predicted from our synthetic spectra and our model of the resolving power of the spectrograph, yielding a best fit for $\vsini = 0$. 
We stress that rotational broadening is significantly smaller than the convective line broadening, hence a slight overestimate of the latter would require a much greater reduction in the former to compensate and therefore possibly implying a vanishing (or unphysical negative) rotational broadening. Still we argue that it is very reassuring that our mean estimates of the stellar rotation velocities at the level of $1$\,km\,s$^{-1}$ are consistent with expectations for old, metal-poor dwarf stars, which to our knowledge has not been reliably achieved previously.

\begin{figure*}
    \centering
    \includegraphics[width=\textwidth]{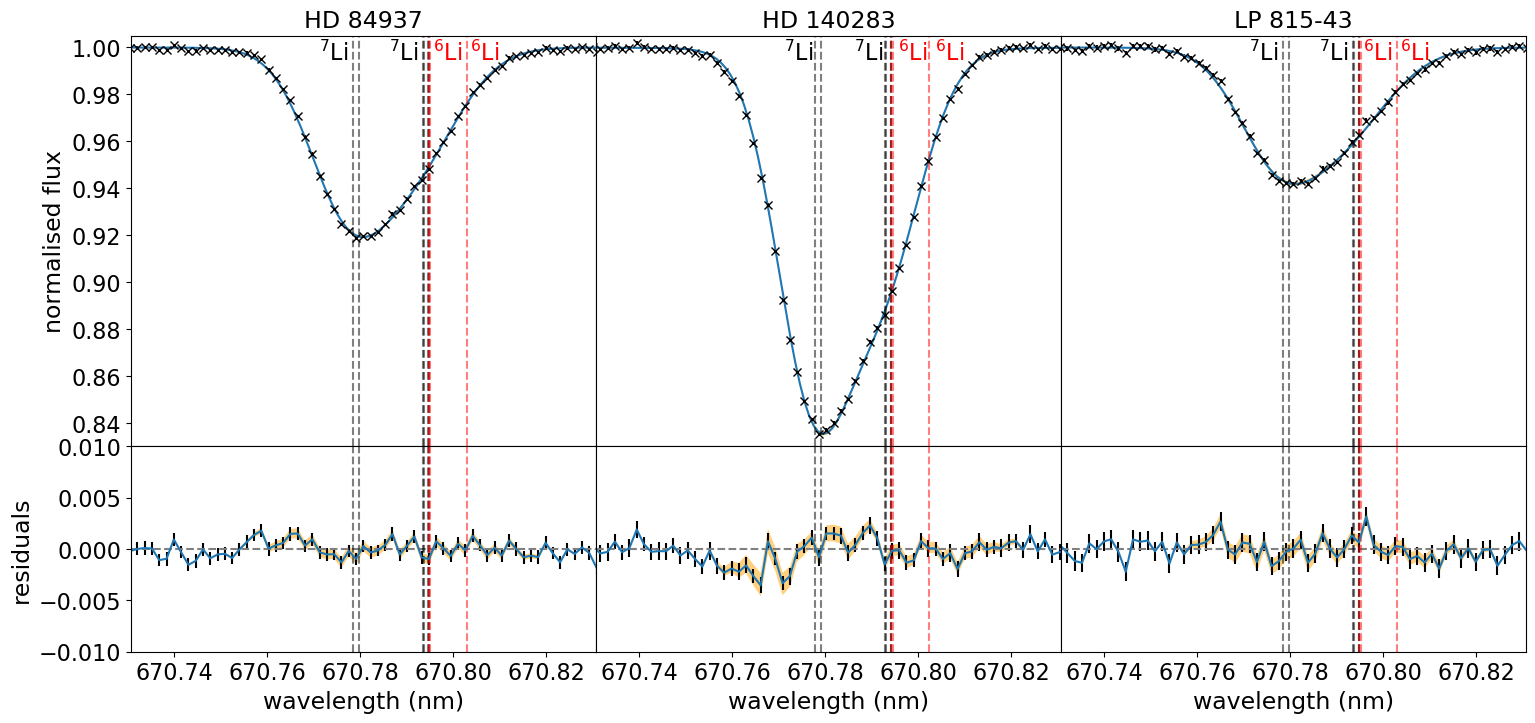}
    \caption{Observed Li with fitted synthetic spectra and residuals in air wavelengths. The wavelengths of isotopic splitting is shown in black ($^7$Li) and red ($^6$Li) dotted lines. The shaded orange region is $99\%$ of the best fit: K constrained for HD 84937 and HD 140283, 4 parameter Li fit for LP 815-43.}
    \label{fig:li_fits}
\end{figure*}

Fig.~\ref{fig:li_fits} shows the best fit for the \liline\ line for all three stars. The orange shaded region contains 99\% of the fits using the fitting method with the smallest errors: the K-constrained fit for HD 84937 and HD 140283, and the Li-only fit for LP 815-43. The synthetic line profiles follow the observations remarkably well, especially considering the extremely high S/N of the observed data. The shaded 99\% region is comparable to the error bars indicating that MCMC is sampling the true distribution of the measured parameters. 

\section{Discussion}
\label{sec:diss}
We employ 3D NLTE synthetic spectra of \liI, \feI, and \kI\ to measure $\ratio$ in three Spite plateau stars: HD 84937, HD 140283, and LP 815-43. We measure 2$\sigma$ upper limits to $\ratio$ of 0.7\,\%, 0.6\,\%, 1.7\,\% respectively for HD 84937, HD 140283, and LP 815-43. Comparing to literature results, for HD 84937, 1D LTE measurements of $\ratio$ typically measure a 2$\sigma$ detection at the 5\% level regardless of use of calibration lines \citep{smith93, smith98, hobbs94, cayrel99}. \citet{steffen12} also measures a similar $\ratio$ with 3D NLTE models. However, \citet{lind13} measures no $\lisi$ with 3D NLTE models regardless of calibration lines. For HD 140283, regardless of use of calibration lines or modelling method there is no detection of $\lisi$ \citep{hobbs94, asplund06, steffen12, lind13}. LP 815-43 had a detection of $0.046 \pm 0.022$ (1D NLTE) \citep{asplund06}. Our results are consistent with those of \citet{lind13} who find no significant detection of $\lisi$. Our results are robust in light of systematics, as we find consistent results regardless of method. 
For completeness, we also apply our analysis to the KECK/HIRES spectra \citep{vogt94} from \citet{lind13}, and find similar results to theirs, as discussed further in App.~\ref{sec:HIRES}.

Fig.~\ref{fig:ellipse} shows a detection of stellar rotation using K calibration lines (K-constrained) in HD 84937 and fitting only on the Li line (Li-only) in HD 140283. However, we caution the reader in these stellar rotation detections as they vary depending on the set of calibration lines chosen. This is further shown in Fig.~\ref{fig:fe_dist}, where the spread in Fe fits for stellar rotation is much larger than individual uncertainties, so some Fe lines detect stellar rotation, whilst others do not. The rotational velocity has a slight effect on $\ratio$, in particular, a lower rotational velocity forces a higher $\ratio$. However, the measured $\ratio$ is still consistent within error bars regardless of the detection of stellar rotation or not.

The fits using Fe calibration lines (Fe-constrained) have larger errors due to the larger uncertainty in the rotational and radial velocity measurements for Fe compared to other fitting methods. Part of this larger uncertainty is due to the lower S/N of LP 815-43 compared to HD 84937 and HD 140283. In addition, the larger uncertainty in $\tvr$ is likely caused by the typical $\sim$25 \ms\ wavelength uncertainties involved in determining the transition wavelengths for Fe. It is worth noting that the uncertainty we find on $\tvr$ through Fe lines is $\times$10 larger than the uncertainty in Fe atomic data, indicating that other error sources are likely involved. A similarly larger error than expected is found on radial velocities by \citet{hernandez_solar_2020}. On the other hand, this uncertainty is only 0.06 \ms\ for K. In addition to the uncertainties involved in determining the transition wavelengths for Fe, the wavelength solution exhibits systematic errors of the order of 20 \ms\ \citep{schmidt_fundamental_2020} which is comparable to our fitting error. 
As the uncertainties in radial velocity are driving the uncertainties in $\ratio$, we strongly encourage an improvement in the reference Fe lines to enable better radial velocity measurements using Fe lines.

We find a residual interference pattern of unknown origin in the high S/N ESPRESSO spectra, also seen in other recent works, see \citet{allart_wasp-127b_2020, casasayas-barris_atmosphere_2021}. Due to the small amplitude of the interference pattern, it is only visible in spectra with S/N $>$ 1000. However, this pattern will affect all results which rely on accurate flux measurements. We remove this interference pattern only for the Li line through use of Gaussian processes, as this technique requires a large line-free region. The effect of this residual interference pattern is $<$1\,\% on our $\ratio$ measurement, which is not enough to affect our conclusions. 

\bba{\citet{lind13} made a conservative estimate of the uncertainty in 3D LTE inferred $\ratio$ ratio of 0.4--0.9\,\%, depending on the number of free parameters, driven by a an assumed error in $\teff$ of 100\,K. We cannot straightforwardly translate this into an appropriate systematic uncertainly of our measurements, due to the many differences in analysis technique with respect to \citet{lind13}, but we estimate that with our $\teff$ uncertainties we cannot rule out an error contribution of order 0.5\,\%. We note that this is similar to our quoted error bars for $\ratio$ for HD 84937 and HD 140283, however, this would not alter our claim of not detecting $\lisi$ in this work. Had $\teff$ been significantly underestimated for a particular star, then this would cause a lack of thermal and convective broadening, which would result in a large $\vsini$ measurement. Conversely, an overestimate of $\teff$ would result in significantly too narrow synthetic profiles. We do not see evidence of either case in our analysis.}

Our synthetic spectra fit well to the high S/N ESPRESSO spectra overall. In particular, the weaker Fe lines also produce fits with residuals within flux error. 
Our results put strong constraints on the presence of $\lisi$ in three well-studied Spite Plateau stars\bba{, which constrains exotic BBN theories that predict a significant amount of primordial $\lisi$ \citep[e.g.][]{luo21}.}
We find that there is no longer any convincing evidence of $\lisi$ in very metal-poor stars.

\section*{Acknowledgements}
\bba{We thank the anonymous referee for their many useful suggestions that improved the quality of this paper.} We thank Michael J. Ireland and Yuan-Sen Ting for useful discussions that improved the quality of this work. 
This research was supported by the Australian Research Council Centre of Excellence for All Sky Astrophysics in 3 Dimensions (ASTRO 3D), through project number CE170100013. 
This work was supported by computational resources provided by the Australian Government through the National Computational Infrastructure (NCI) under the National Computational Merit Allocation Scheme and the ANU Merit Allocation Scheme \bb{(project y89)}.
MA gratefully acknowledges funding from the Australian Research Council (grants FL110100012 and DP150100250).
KL acknowledges funds from the European Research Council (ERC) under the European Union’s Horizon 2020 research and innovation programme (Grant agreement No. 852977)
Based on observations made with ESO Telescopes at the La Silla Paranal Observatory under programme ID 0103.D-0616(A). 

\section*{Data availability}
The observations are publicly available in the ESO Science Archive Facility under programme ID 0103.D-0616(A). Reduced spectra and simulation results will be shared upon request. 

\bibliographystyle{mnras}
\bibliography{ref}

\appendix

\section{Gaussian process test}
\label{sec:GP_test}
\begin{figure*}
    \centering
    \includegraphics[width=\textwidth]{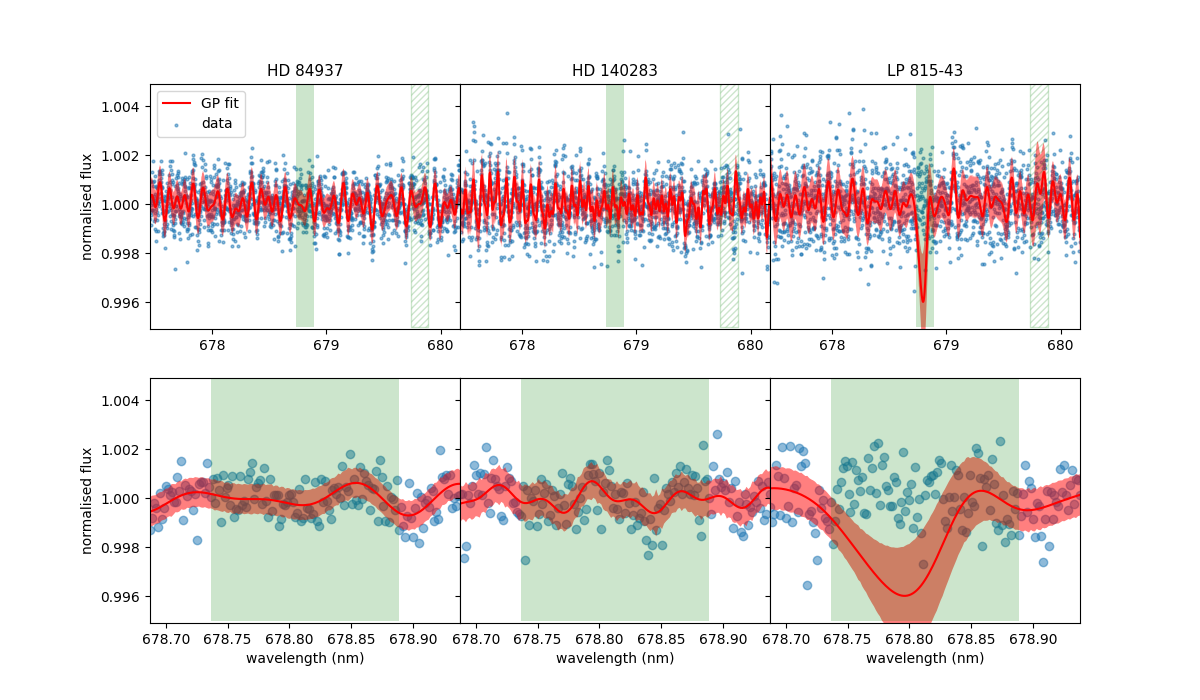}
    \caption{The residual interference pattern at $\sim 680$\,nm in HD 84937 (left), HD 140283 (middle), and LP 815 (right). Blue shows the stacked observation with the residual interference pattern, red shows the Gaussian process fit to this residual interference pattern, and shaded red is the flux error and Gaussian process error summed in quadrature. Green shows the region which was not considered in the fit. The bottom panel shows the green masked region at 679\,nm.}
    \label{fig:gp_test}
\end{figure*}

We test our Gaussian process on a line-free region centered at $679$\,nm with the same masking as the \liline, shown in Fig.~\ref{fig:gp_test}. Qualitatively, the Gaussian process fit to the masked out region is good except for LP 815-43. Not only was the interference pattern predicted poorly in LP 815-43, but also the associated Gaussian process error was larger, indicating \bba{an uncertain} prediction. Quantitatively, the measured S/N (given by the MAD) for the green masked region at 679\,nm is 1305, 1087, and 901 respectively for HD 84937, HD 140283, LP 815-43. After fitting the Gaussian process and removing this fit from the data, the S/N is 1314, 1258, and 443 respectively. The visibly poor fit for LP 815-43 is reflected in the S/N where LP 815-43 decreased in S/N after removal. Therefore, we do not remove the Gaussian process fit from the observed spectra of LP 815-43. 

\section{Calibration line fits}
\label{sec:fe_fits}
\begin{figure*}
    \centering
    \includegraphics[width=0.27\textwidth]{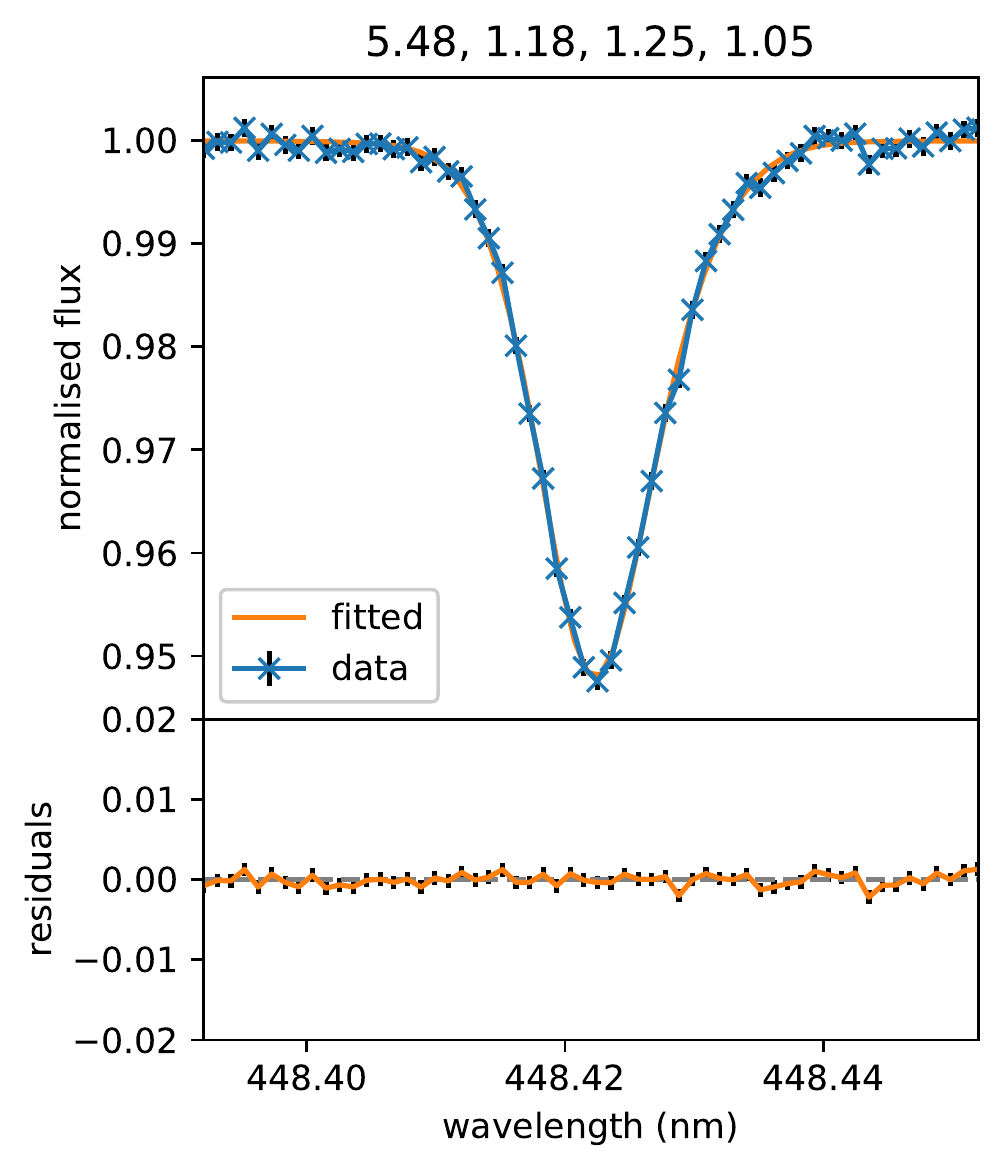}
    \includegraphics[width=0.27\textwidth]{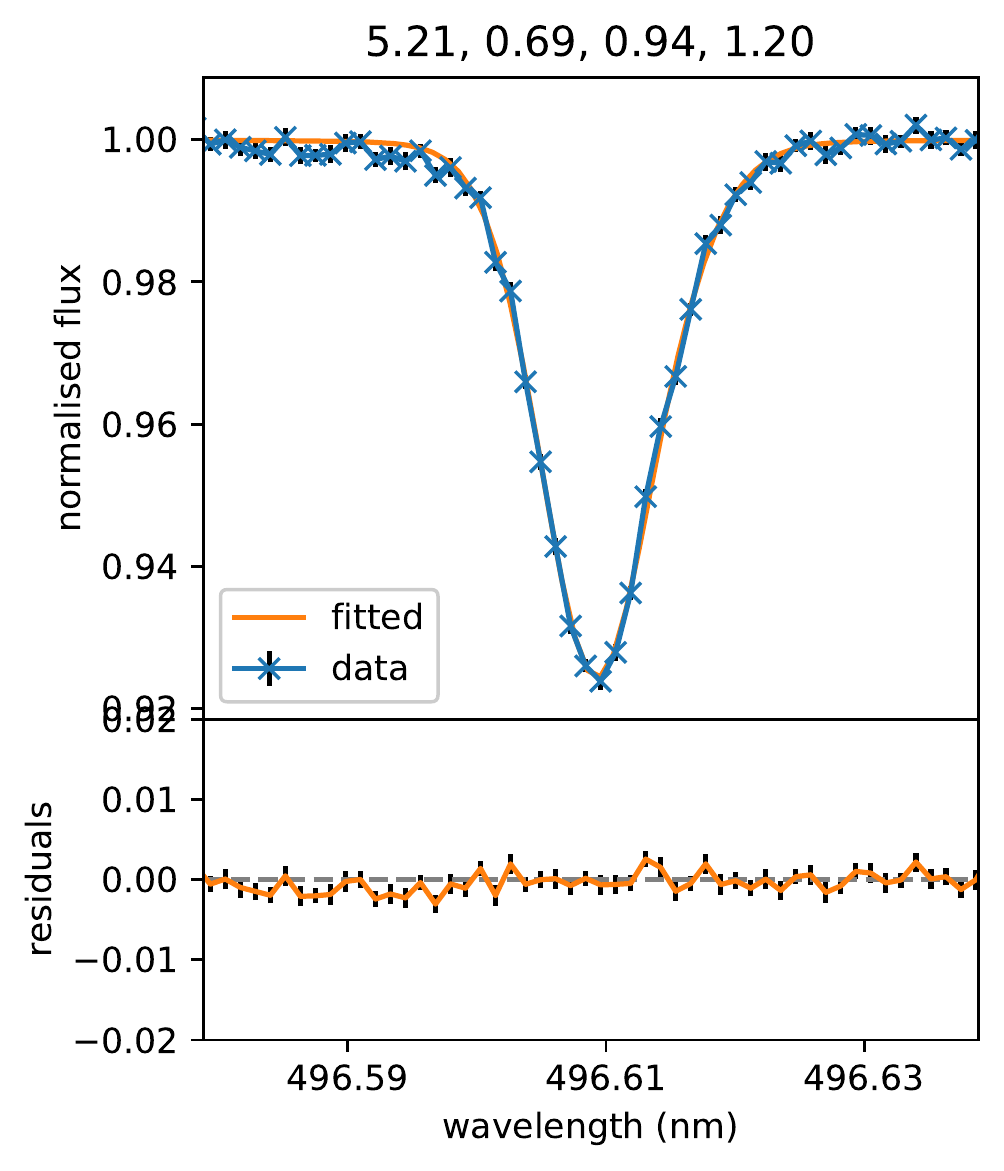}
    \includegraphics[width=0.27\textwidth]{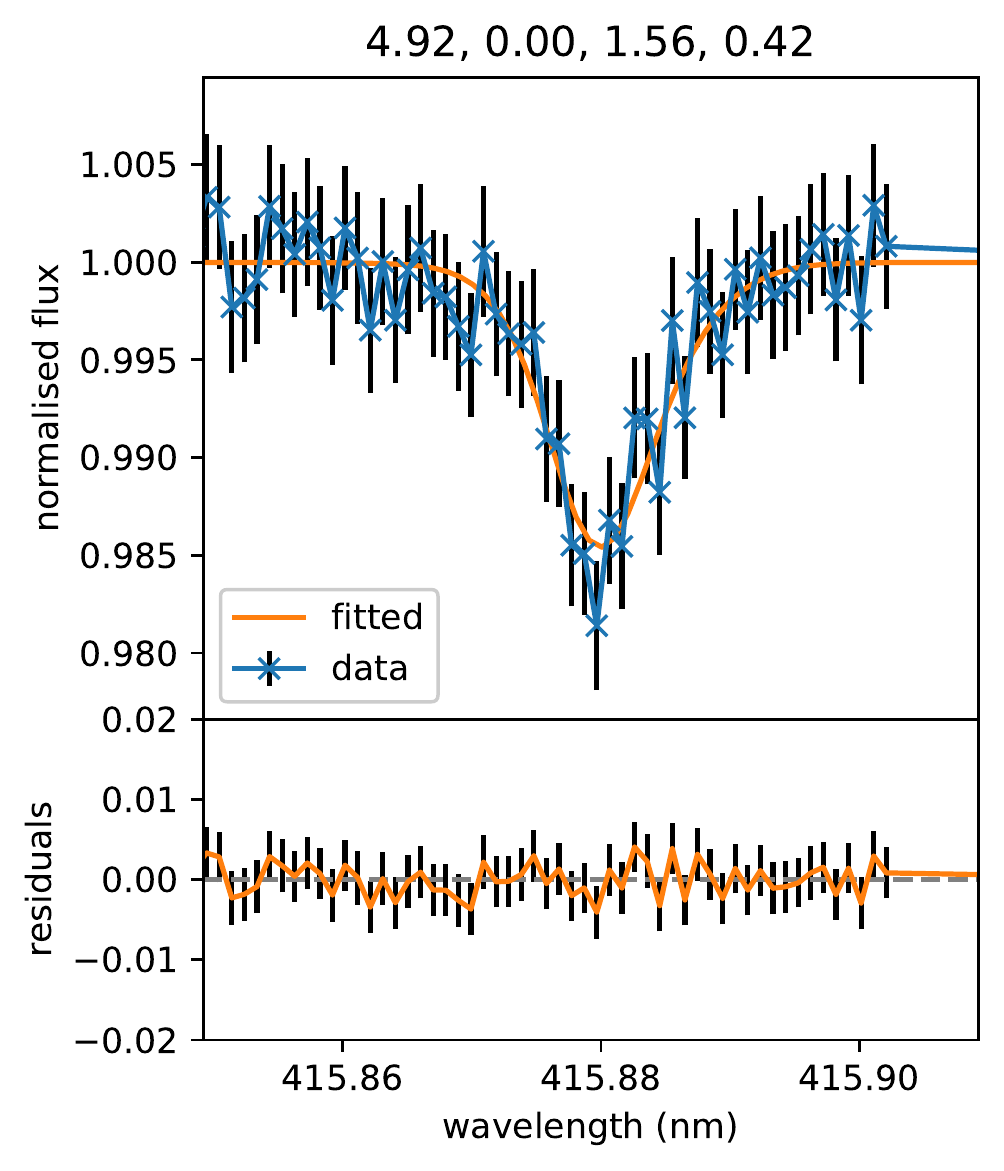}
    \includegraphics[width=0.27\textwidth]{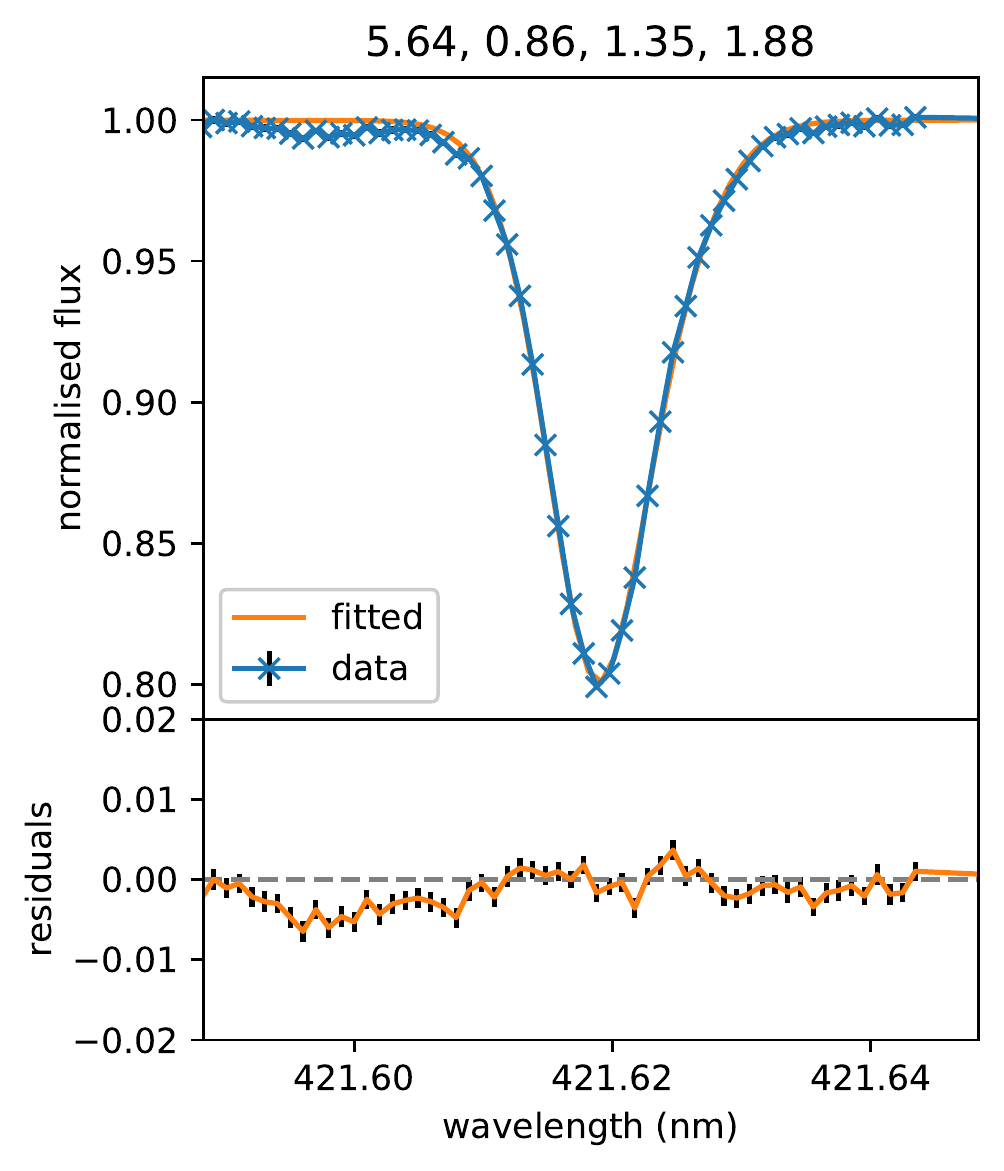}
    \includegraphics[width=0.27\textwidth]{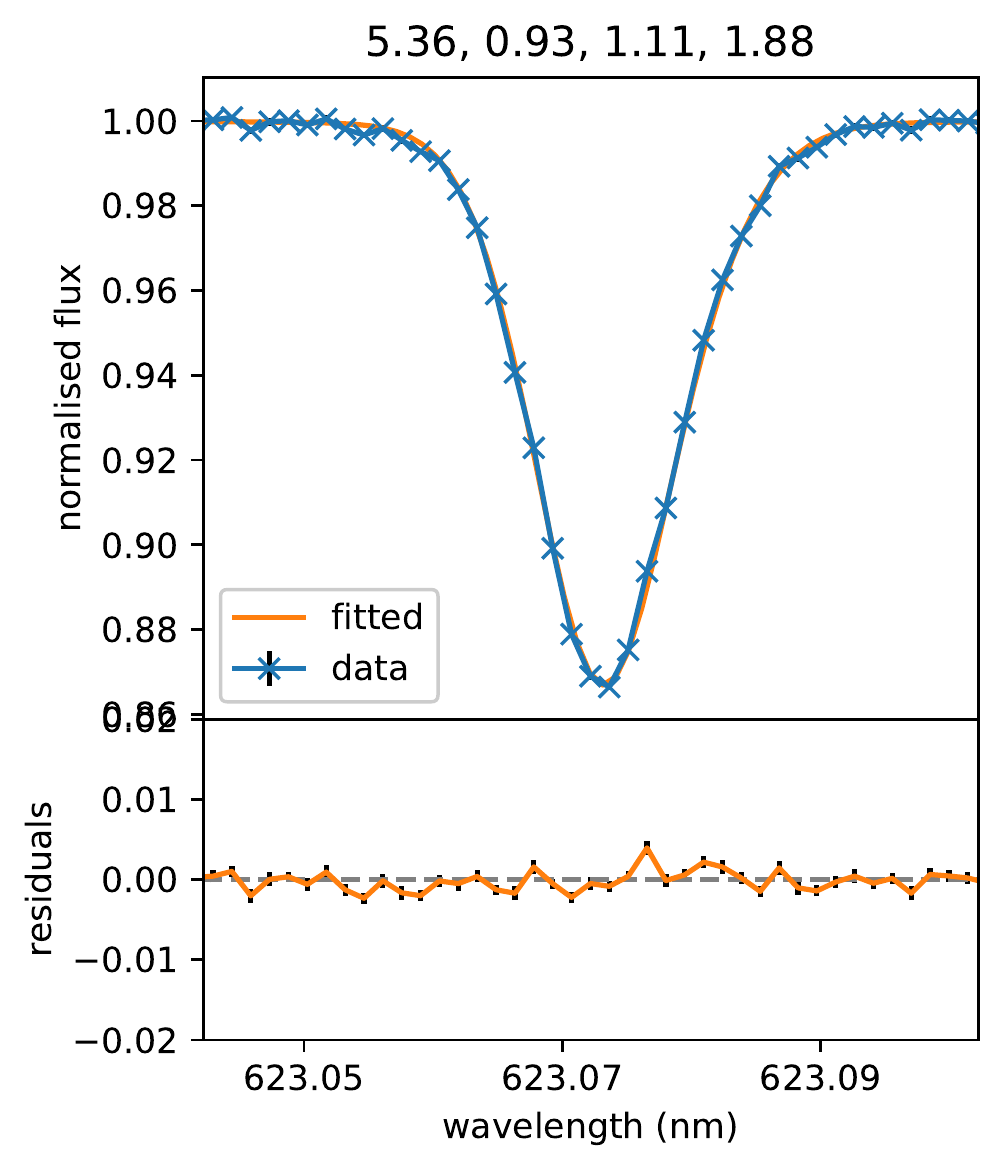}
    \includegraphics[width=0.27\textwidth]{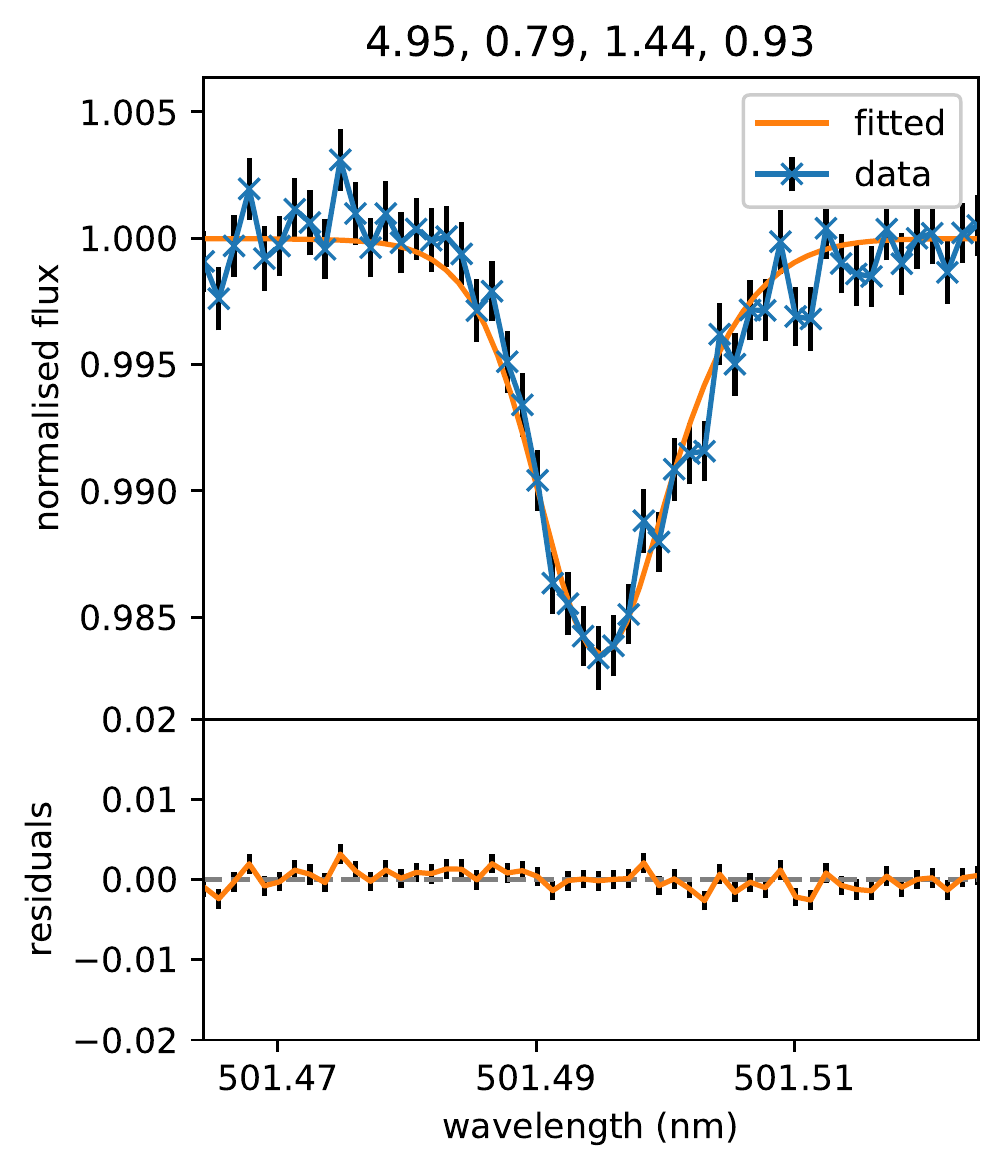}
    \includegraphics[width=0.27\textwidth]{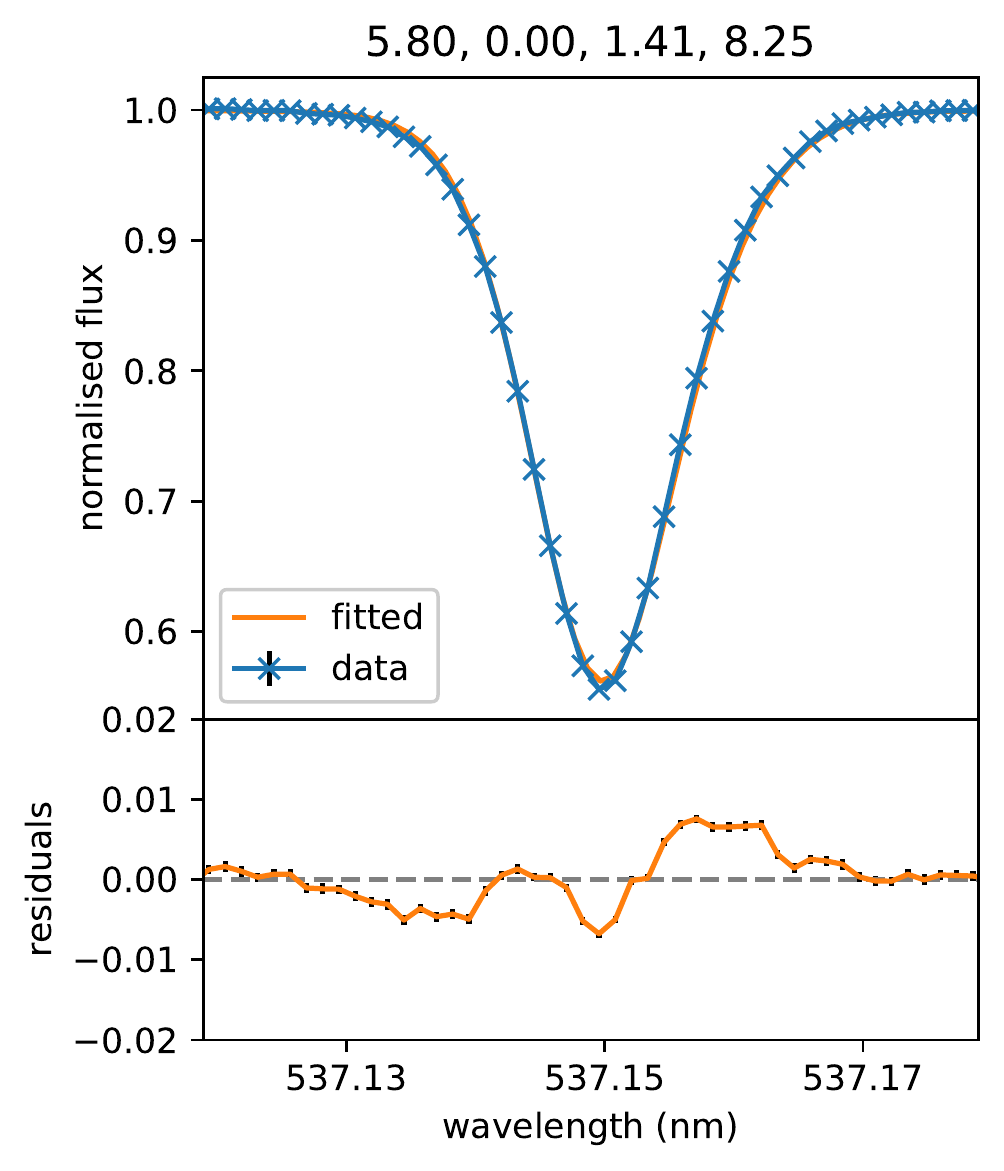}
    \includegraphics[width=0.27\textwidth]{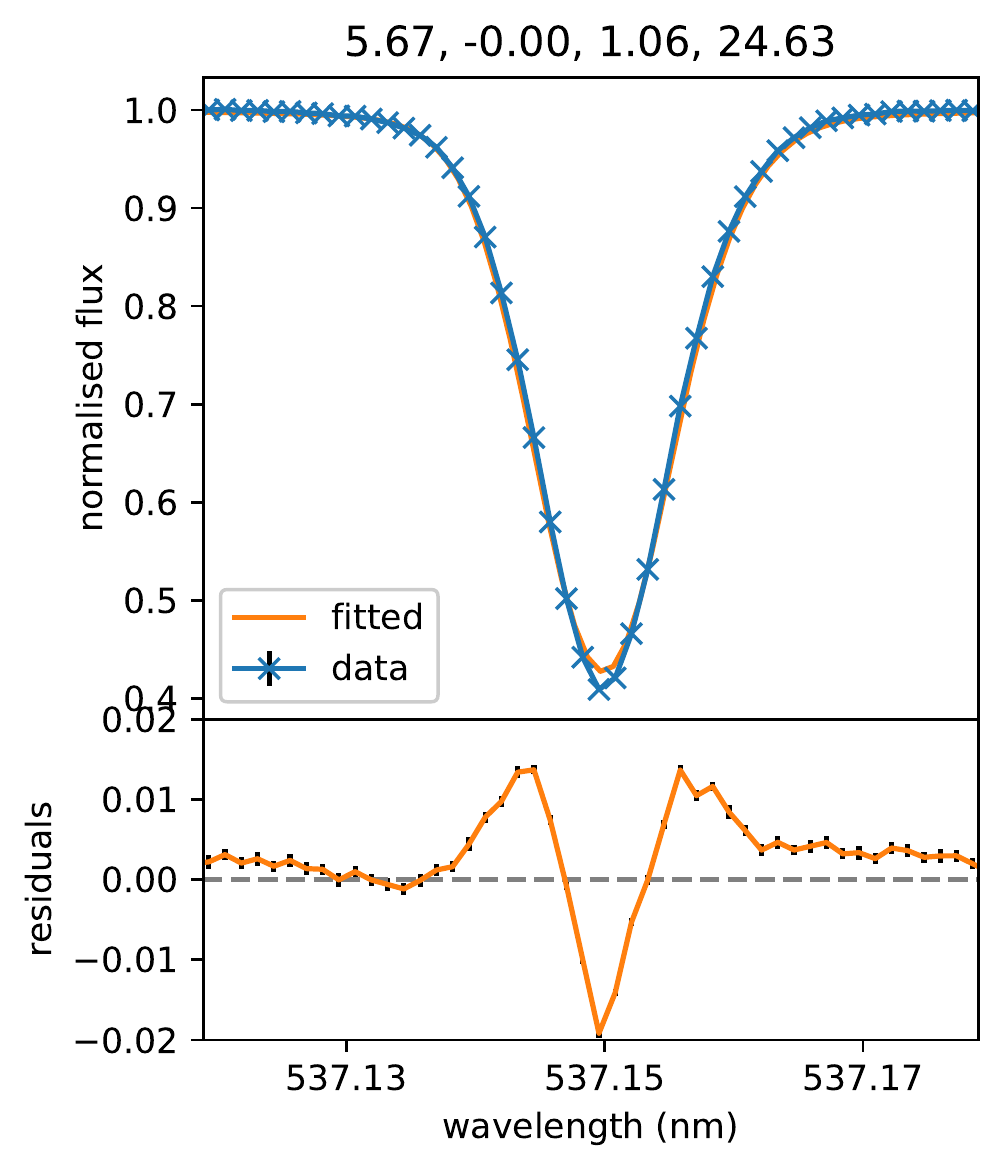}
    \includegraphics[width=0.27\textwidth]{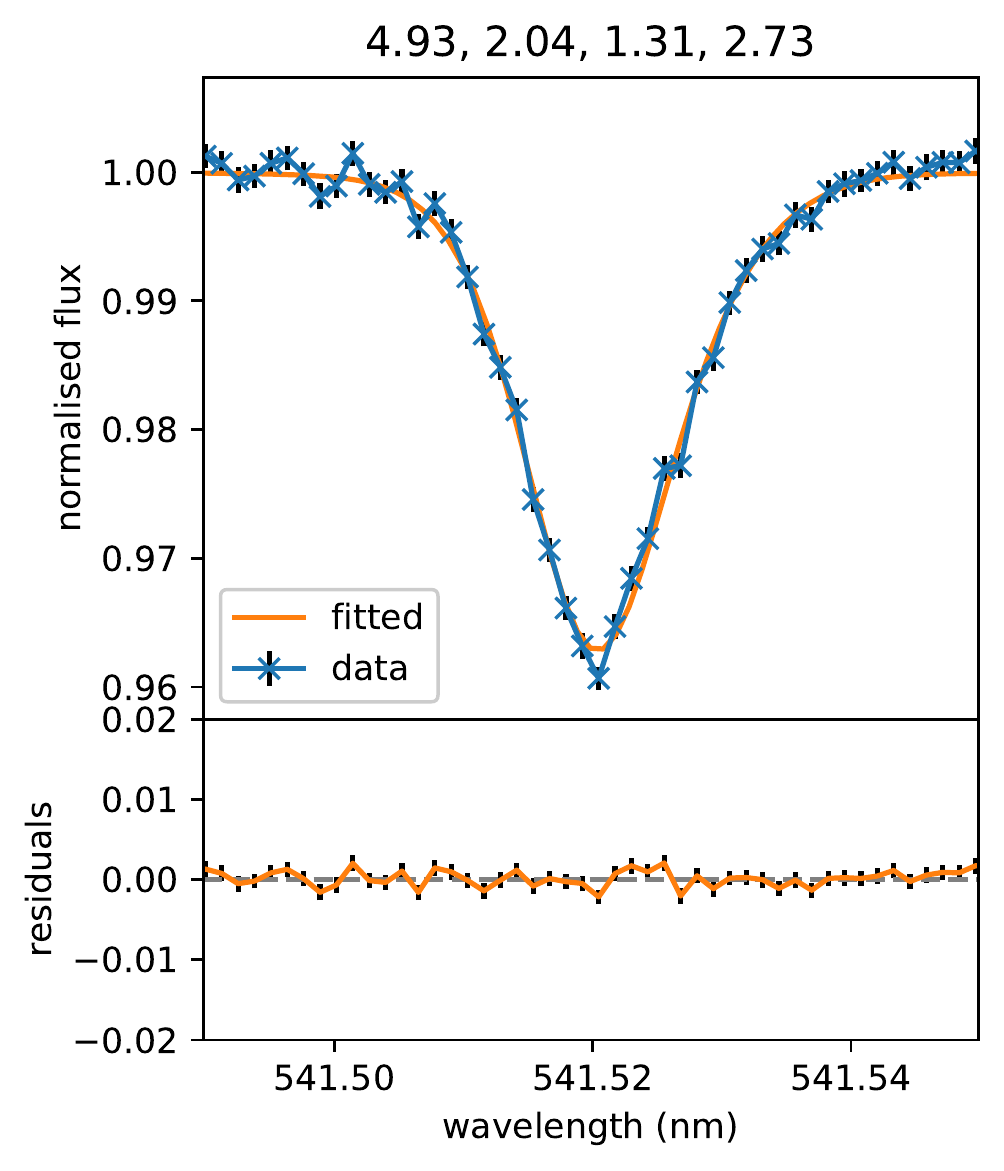}
    \includegraphics[width=0.27\textwidth]{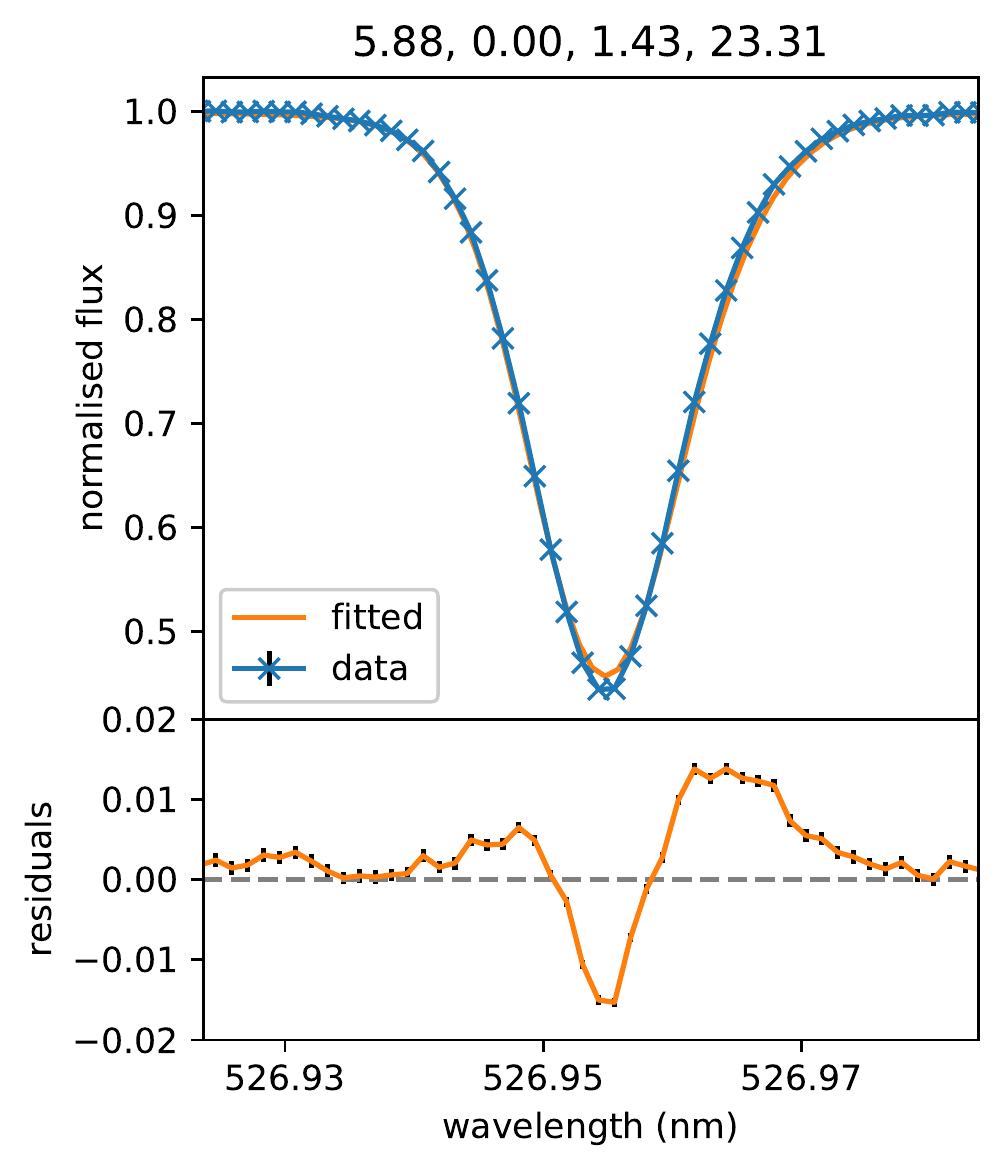}
    \includegraphics[width=0.27\textwidth]{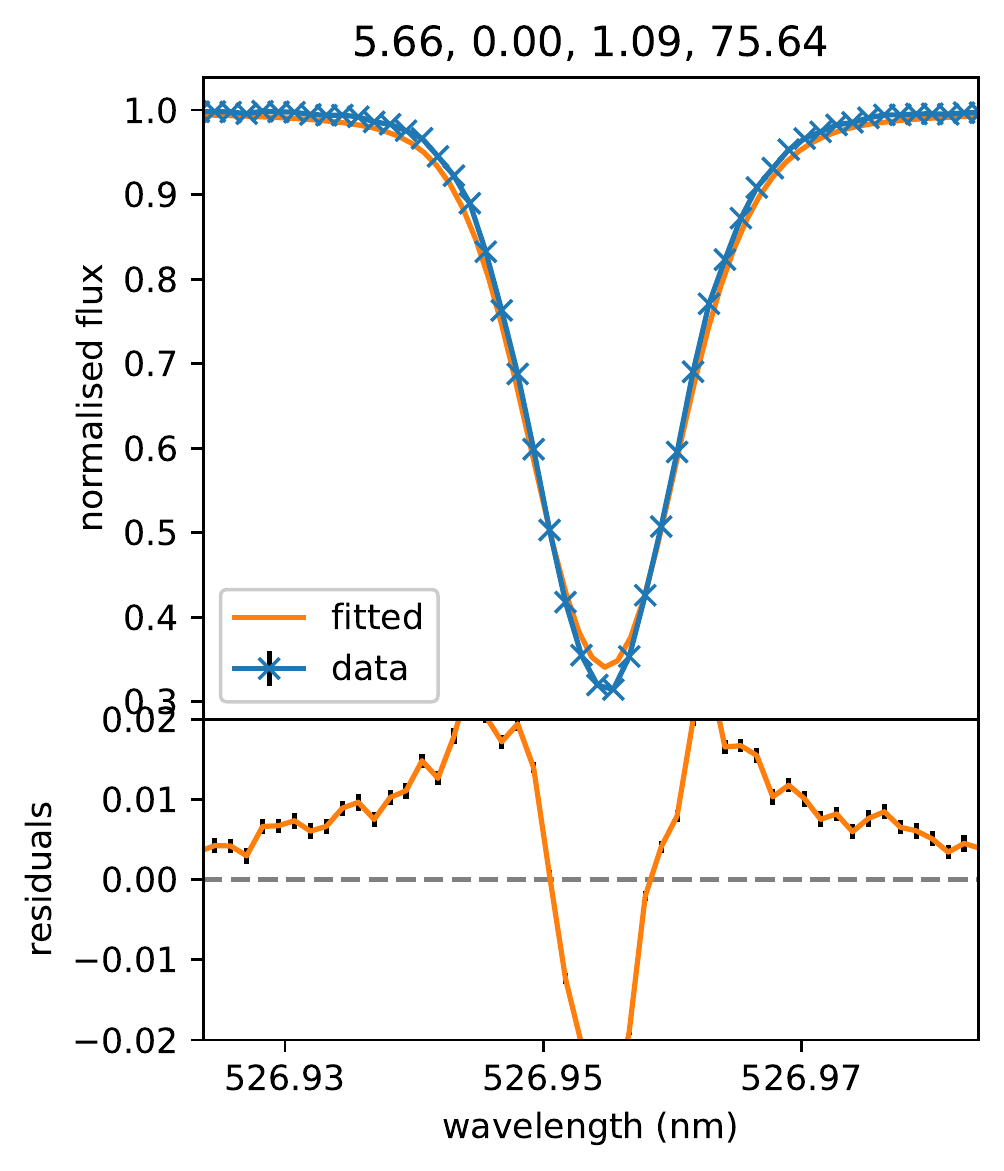}
    \includegraphics[width=0.27\textwidth]{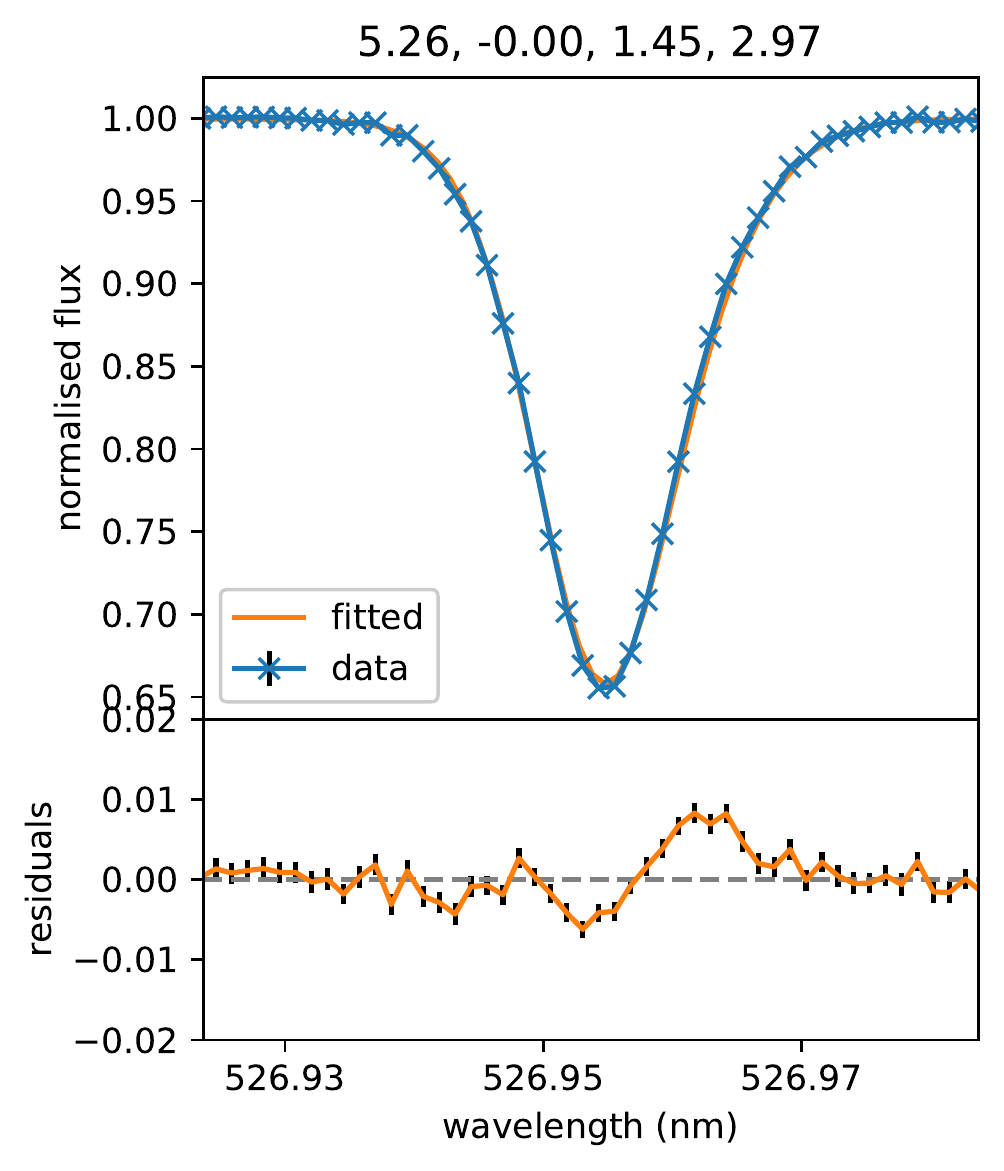}
    \caption{The observed (blue) line profile, fitted (orange) line profile, and residuals for some \feI\ lines for the three observed stars HD 84937 (left column), HD 140283 (middle column), and LP 815-43 (right column). The top row shows the fit with the smallest $\chi^2_{red}$. The second row from the top shows the fit with a medium $\chi^2_{red}$. The bottom two rows show the fits for the two largest $\chi^2_{red}$. The fitted A(Fe), $\vsini$, $\tvr$, and $\chi^2_{red}$ are respectively labelled above each fit.}
    \label{fig:fe_fits}
\end{figure*}

We fit 47 \feI\ lines for each observed star. The best, medium, and two worst fits are shown in Fig.~\ref{fig:fe_fits}. The worst fits usually measure $\vsini=0$. 

The best fits for the \kline\ line and all 47 \feI\ lines for HD 84937, HD 140283, and LP 815-43 are tabulated in Table.~\ref{tab:fe_fits}. In general, we find worse fitting results for the more saturated lines.

\begin{table*}
    \centering
    \caption{The A(X) (dex) where X represents some element, $\vsini$ (\kms), and $\vr$ (\kms) that give the maximum likelihood fits for HD 84937, HD 140283, and LP 815-43. The fitted EW (pm) and $\chi^2_{red}$ computed with the maximum likelihood fits are also included.}
\begin{tabular}{c|ccccc|ccccc|ccccc}
& \multicolumn{5}{|c|}{HD 84937} &\multicolumn{5}{|c|}{HD 140283} &\multicolumn{5}{|c|}{LP 815-43} \\
$\lambda$ & A(X) & $\svsini$ & $\vr$ & EW & $\chi^2_{red}$ & A(X) & $\svsini$ & $\vr$ & EW & $\chi^2_{red}$ & A(X) & $\svsini$ & $\vr$ & EW & $\chi^2_{red}$ \\
\hline
\hline 
\multicolumn{16}{c}\kI \\
769.90 & 3.31 & 1.49 & 1.26 & 1.59 & 2.42 & 2.80 & 0.40 & 1.05 & 1.25 & 1.33 \\
\\
\multicolumn{16}{c}\feI \\
414.77 & 5.54 & 0.93 & 1.38 & 1.62 & 1.38 & 5.28 & 0.16 & 1.03 & 2.30 & 1.59 & 4.99 & 0.00 & 1.26 & 0.39 & 0.80 \\
415.88 & 5.49 & 1.08 & 1.25 & 0.75 & 1.25 & 5.19 & 0.90 & 0.88 & 0.77 & 1.50 & 4.92 & 0.00 & 1.56 & 0.16 & 0.42 \\
421.62 & 5.64 & 0.86 & 1.36 & 2.17 & 1.88 & 5.46 & 0.00 & 1.12 & 3.76 & 2.37 & 5.09 & 0.00 & 1.37 & 0.53 & 0.60 \\
428.24 & 5.60 & 1.25 & 1.39 & 4.25 & 2.26 & 5.44 & 1.34 & 1.07 & 4.81 & 6.33 & 5.00 & 0.00 & 1.43 & 1.51 & 0.85 \\
437.59 & 5.65 & 0.00 & 1.40 & 3.67 & 3.84 & 5.55 & 0.00 & 1.09 & 5.58 & 14.04 & 5.09 & 0.00 & 1.35 & 1.10 & 1.04 \\
446.17 & 5.62 & 0.57 & 1.37 & 2.48 & 3.58 & 5.46 & 0.00 & 1.09 & 4.16 & 3.29 & 5.10 & 1.78 & 1.41 & 0.68 & 0.72 \\
448.42 & 5.48 & 1.18 & 1.24 & 0.62 & 1.05 & 5.16 & 0.00 & 0.92 & 0.60 & 1.37 & 4.93 & 0.00 & 1.32 & 0.14 & 0.83 \\
449.46 & 5.61 & 0.89 & 1.38 & 2.90 & 1.79 & 5.35 & 0.76 & 1.07 & 3.46 & 1.80 & 5.04 & 0.00 & 1.33 & 0.83 & 0.57 \\
452.86 & 5.72 & 1.62 & 1.44 & 4.50 & 7.56 & 5.50 & 0.85 & 1.10 & 5.10 & 5.16 & 5.11 & 1.35 & 1.44 & 1.59 & 0.85 \\
490.33 & 5.49 & 0.91 & 1.36 & 1.39 & 1.52 & 5.21 & 0.74 & 1.01 & 1.63 & 1.24 & 4.95 & 1.13 & 1.28 & 0.35 & 1.01 \\
491.90 & 5.56 & 1.22 & 1.32 & 3.60 & 2.42 & 5.29 & 0.88 & 1.01 & 3.95 & 2.38 & 4.98 & 0.64 & 1.37 & 1.14 & 0.67 \\
493.88 & 5.51 & 0.00 & 1.33 & 1.02 & 1.87 & 5.22 & 0.00 & 1.00 & 1.20 & 1.25 & 4.92 & 0.00 & 1.14 & 0.21 & 1.05 \\
496.61 & 5.51 & 0.59 & 1.28 & 0.81 & 1.26 & 5.21 & 0.69 & 0.94 & 0.87 & 1.20 & 4.99 & 0.00 & 1.34 & 0.20 & 0.76 \\
500.19 & 5.50 & 1.15 & 1.37 & 1.39 & 2.19 & 5.20 & 1.16 & 1.02 & 1.33 & 1.46 & 4.97 & 0.91 & 1.46 & 0.37 & 0.74 \\
500.61 & 5.52 & 1.24 & 1.35 & 2.48 & 1.40 & 5.25 & 0.76 & 1.03 & 2.81 & 2.05 & 4.96 & 1.21 & 1.28 & 0.68 & 0.90 \\
501.21 & 5.63 & 1.25 & 1.45 & 2.07 & 1.52 & 5.42 & 0.37 & 1.13 & 3.24 & 2.34 & 5.09 & 0.00 & 1.44 & 0.52 & 0.70 \\
501.49 & 5.49 & 1.51 & 1.27 & 0.90 & 1.44 & 5.19 & 1.14 & 0.97 & 0.85 & 1.41 & 4.95 & 0.79 & 1.44 & 0.22 & 0.93 \\
504.98 & 5.55 & 0.36 & 1.38 & 1.74 & 2.20 & 5.29 & 0.00 & 1.08 & 2.20 & 1.44 & 5.01 & 1.25 & 1.54 & 0.44 & 0.86 \\
505.16 & 5.61 & 0.64 & 1.37 & 1.39 & 1.85 & 5.38 & 0.00 & 1.06 & 2.32 & 2.03 & 5.05 & 0.00 & 1.21 & 0.31 & 1.28 \\
511.04 & 5.69 & 1.99 & 1.16 & 1.29 & 2.09 & 5.45 & 0.87 & 0.97 & 2.47 & 2.38 & 5.19 & 3.00 & 1.01 & 0.31 & 0.64 \\
513.37 & 5.46 & 1.30 & 1.39 & 1.63 & 1.17 & 5.16 & 1.06 & 1.04 & 1.47 & 1.26 & 4.93 & 1.51 & 1.34 & 0.45 & 0.76 \\
517.16 & 5.56 & 0.77 & 1.42 & 3.12 & 1.92 & 5.36 & 0.00 & 1.10 & 4.16 & 3.78 & 5.01 & 0.00 & 1.39 & 0.92 & 0.83 \\
519.15 & 5.52 & 1.58 & 1.40 & 2.06 & 2.71 & 5.21 & 0.72 & 1.05 & 2.24 & 1.49 & 4.95 & 1.30 & 1.29 & 0.54 & 0.85 \\
519.23 & 5.52 & 1.34 & 1.40 & 2.68 & 2.72 & 5.24 & 0.27 & 1.08 & 2.91 & 1.78 & 4.95 & 0.79 & 1.38 & 0.75 & 0.88 \\
519.49 & 5.53 & 0.83 & 1.40 & 1.71 & 2.10 & 5.29 & 0.35 & 1.09 & 2.46 & 1.96 & 5.01 & 0.69 & 1.41 & 0.43 & 0.72 \\
521.52 & 5.44 & 1.35 & 1.31 & 0.70 & 1.22 & 5.14 & 1.22 & 0.94 & 0.76 & 1.22 & 4.92 & 1.51 & 1.48 & 0.17 & 0.67 \\
521.63 & 5.51 & 0.84 & 1.36 & 1.37 & 1.47 & 5.26 & 0.61 & 1.07 & 1.99 & 1.63 & 4.98 & 0.00 & 1.14 & 0.33 & 0.67 \\
526.95 & 5.88 & 0.00 & 1.43 & 8.61 & 24.00 & 5.66 & 0.00 & 1.07 & 10.14 & 76.32 & 5.26 & 0.00 & 1.45 & 5.08 & 2.98 \\
528.18 & 5.48 & 0.26 & 1.37 & 1.19 & 1.45 & 5.19 & 0.00 & 1.05 & 1.34 & 1.22 & 4.86 & 0.00 & 1.47 & 0.24 & 2.71 \\
532.42 & 5.53 & 1.50 & 1.36 & 3.18 & 1.70 & 5.25 & 1.25 & 1.02 & 3.38 & 1.82 & 4.97 & 1.37 & 1.37 & 1.00 & 1.73 \\
533.99 & 5.48 & 0.79 & 1.31 & 1.20 & 1.70 & 5.19 & 1.12 & 0.99 & 1.31 & 1.24 & 4.94 & 1.65 & 1.35 & 0.29 & 1.12 \\
534.10 & 5.62 & 1.23 & 1.39 & 2.09 & 2.24 & 5.39 & 0.67 & 1.09 & 2.91 & 2.42 & 5.09 & 0.00 & 1.29 & 0.55 & 1.24 \\
537.00 & 5.44 & 1.16 & 1.31 & 1.60 & 1.69 & 5.13 & 1.04 & 1.00 & 1.39 & 1.41 & 4.91 & 0.00 & 1.34 & 0.45 & 1.31 \\
537.15 & 5.80 & 0.00 & 1.42 & 6.61 & 9.04 & 5.67 & 0.00 & 1.06 & 8.12 & 25.41 & 5.14 & 0.00 & 1.43 & 2.95 & 2.20 \\
539.32 & 5.50 & 1.23 & 1.33 & 1.09 & 1.35 & 5.20 & 1.17 & 1.00 & 1.18 & 1.34 & 4.97 & 2.10 & 1.50 & 0.27 & 1.13 \\
539.71 & 5.72 & 0.00 & 1.45 & 4.97 & 5.36 & 5.62 & 0.00 & 1.09 & 6.60 & 13.08 & 5.10 & 0.00 & 1.38 & 1.74 & 1.48 \\
540.58 & 5.72 & 0.00 & 1.44 & 5.33 & 5.88 & 5.61 & 0.00 & 1.09 & 6.87 & 15.23 & 5.09 & 0.00 & 1.41 & 1.96 & 1.99 \\
541.52 & 5.44 & 1.75 & 1.31 & 1.88 & 3.20 & 5.13 & 1.71 & 0.93 & 1.65 & 1.45 & 4.93 & 2.04 & 1.31 & 0.57 & 2.73 \\
542.97 & 5.75 & 0.00 & 1.43 & 5.43 & 7.90 & 5.65 & 0.00 & 1.08 & 7.04 & 13.20 & 5.12 & 0.00 & 1.40 & 2.04 & 1.36 \\
543.45 & 5.67 & 0.00 & 1.45 & 3.83 & 5.21 & 5.53 & 0.00 & 1.09 & 5.28 & 9.16 & 5.08 & 0.00 & 1.38 & 1.14 & 1.49 \\
549.75 & 5.61 & 1.21 & 1.34 & 1.12 & 1.78 & 5.36 & 0.06 & 1.03 & 1.85 & 1.62 & 5.08 & 1.74 & 1.56 & 0.25 & 0.98 \\
550.68 & 5.61 & 1.03 & 1.39 & 1.27 & 1.72 & 5.37 & 0.00 & 1.08 & 2.11 & 2.26 & 5.06 & 0.00 & 1.36 & 0.29 & 1.04 \\
556.96 & 5.49 & 1.34 & 1.34 & 1.17 & 2.64 & 5.21 & 1.46 & 0.98 & 1.27 & 1.60 & 4.91 & 0.00 & 1.14 & 0.26 & 2.31 \\
557.28 & 5.51 & 1.66 & 1.31 & 1.90 & 2.16 & 5.21 & 1.26 & 1.00 & 1.98 & 2.60 & 4.96 & 1.81 & 1.29 & 0.50 & 1.26 \\
558.68 & 5.52 & 1.37 & 1.34 & 2.61 & 1.52 & 5.24 & 1.11 & 0.98 & 2.75 & 1.93 & 4.95 & 0.00 & 1.32 & 0.74 & 1.05 \\
561.56 & 5.52 & 1.60 & 1.33 & 3.34 & 2.20 & 5.24 & 1.30 & 1.01 & 3.48 & 2.65 & 4.96 & 1.37 & 1.38 & 1.07 & 1.06 \\
623.07 & 5.63 & 1.25 & 1.42 & 1.54 & 1.68 & 5.36 & 0.93 & 1.09 & 1.89 & 1.83 & 5.08 & 1.21 & 1.52 & 0.37 & 0.93 \\
\hline
    \end{tabular}
    \label{tab:fe_fits}
\end{table*}

\section{Comparison to HIRES}
\label{sec:HIRES}
\begin{figure*}
    \centering
    \includegraphics[width=\textwidth]{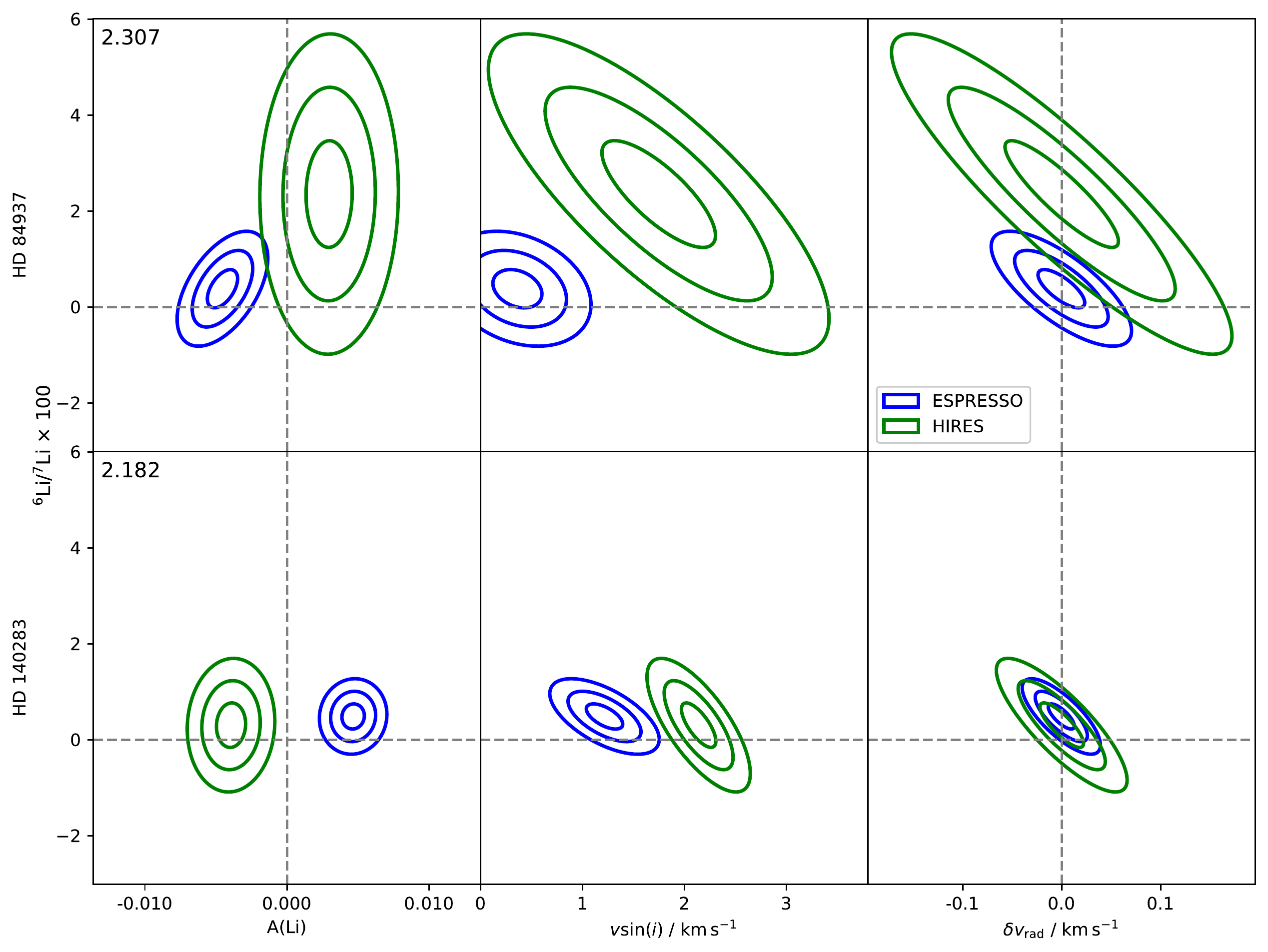}
    \caption{Confidence ellipses containing 1, 2, and 3$\sigma$ of the data of a 4 parameter fit to only the Li line (Li-only) with ESPRESSO (blue) and HIRES data (green).}
    \label{fig:ellipse_lind13}
\end{figure*}

We ran the same Li-only fit on HIRES data from \citet{lind13}. Fig.~\ref{fig:ellipse_lind13} shows the Li-only fit for ESPRESSO and HIRES. The ESPRESSO and HIRES results are mostly in agreement with each other, indicating that the residual interference pattern removal was successful. The HIRES results here are presented with our analysis method, and is in agreement with the analysis presented in \citet{lind13}. The predicted abundance is higher in HD 84937 here than \citet{lind13} as we use a higher temperature stellar atmosphere model. This agreement indicates that the update in statistical method and stellar atmosphere models did not skew the results. We note that the ESPRESSO data reduced the error bars compared to HIRES, in particular for HD 84937. 

\end{document}